\documentclass{jfm}
  \usepackage{amsmath}
  \usepackage{mathabx}

  \usepackage{amssymb}
  \usepackage{gauss}
  \usepackage{bm}
  \usepackage{mwe}
  \usepackage{subfig}
 \usepackage{caption}
 \usepackage{float}
 \usepackage{natbib}
\usepackage{physics}
\usepackage{epstopdf, epsfig}
\usepackage{xcolor}
\usepackage{hyperref}

\usepackage{leftidx}

\usepackage{soul,xcolor}
\setstcolor{red}

\newcommand{\St}{\mbox{\it St}}

\makeatletter
\def\sgn{\mathop{\operator@font sgn}}
\makeatother

\shorttitle{SPOD for analysis of turbulent flows}
\shortauthor{A. Nekkanti and O. T. Schmidt}

\title{Frequency-time analysis, low-rank reconstruction and denoising of turbulent flows using SPOD}
\author{Akhil Nekkanti \and Oliver T. Schmidt\corresp{\email{oschmidt@ucsd.edu}}}

\affiliation{\aff{1}Department of Mechanical and Aerospace
  Engineering, Jacobs School of Engineering, UCSD, 9500 Gilman Drive, La
  Jolla, CA 92093-0411, USA}
   
\begin{document}

\maketitle

\begin{abstract}

Four different applications of spectral proper orthogonal decomposition (SPOD):  low-rank reconstruction, denoising, frequency-time analysis, and prewhitening are demonstrated on large-eddy simulation data of a turbulent jet. SPOD-based low-rank reconstruction can be performed by direct inversion of a truncated SPOD. This spectral inversion problem, however, is ambiguous since SPOD relies on spectral estimation. We demonstrate SPOD-based flow field reconstruction using direct inversion of the SPOD algorithm (frequency-domain approach) and propose an alternative approach based on projection of the time series data onto the modes (time-domain approach). While the SPOD optimally represents the flow in a statistical sense, the time-domain approach seeks an optimal reconstruction of each instantaneous flow field. We further propose a SPOD-based denoising strategy that is based on hard-thresholding of the SPOD eigenvalues. The proposed strategy achieves significant noise reduction while facilitating drastic data compression. In contrast to standard methods of frequency-time analysis such as wavelet transform, a proposed SPOD-based approach yields a spectrogram that characterizes the temporal evolution of spatially coherent flow structures. In the frequency-domain, time-varying expansion coefficients can be obtained by basing the SPOD on a sliding window. This approach, however, is computationally intractable, and an alternative strategy based on convolution in the time-domain is presented. When applied to the turbulent jet data, SPOD-based frequency-time analysis reveals that the intermittent occurrence of large-scale coherent structures is directly associated with high-energy events. This work suggests that the time-domain approach is preferable for low-rank reconstruction of individual snapshots, and the frequency-domain approach for denoising and frequency-time analysis. 

\end{abstract}


\section{Introduction} \label{introduction}

The curse of dimensionality  \citep[see, e.g.,][]{meneveau1992search} in the analysis of large turbulent flow data has led to the development of a number of modal decomposition techniques \citep{holmes2012turbulence}. The primary utilities of these techniques are to extract the essential flow features and to provide a low-dimensional representation of the data. Most of these techniques seek modes that lie in the span of the snapshots that constitute the time-resolved data, and adhere to certain mathematical properties that define the decomposition. The arguably most widely used technique is the proper orthogonal decomposition (POD), introduced by \citet{lumley1967structure,lumley1970stochastic}.  A specific flavor of POD, the computationally inexpensive method of snapshots \citep{sirovich1987turbulence, aubry1991hidden}, decompose the flow field into spatial modes and temporal coefficients. Its modes optimally represent the data in terms of its variance, or energy, and are coherent in space and at zero time lag. Another popular method is the dynamic mode decomposition  \citep[DMD, ][]{schmid2010dynamic}, which is rooted in Koopman theory \citep{rowley2009spectral} and assumes an evolution operator that maps the flow field from one snapshot to its next. The DMD modes are characterized by a single frequency and linear amplification rate. Refer to the reviews by \citet{taira2017modal} and \citet{rowley2017model} for summaries of various modal techniques.    

Spectral proper orthogonal decomposition (SPOD) is the frequency-domain variant of POD and computes modes as estimates of the eigenvectors of the cross-spectral density (CSD) matrix. At each frequency, SPOD yields a set of orthogonal modes, ranked by energy. The mathematical framework underlying SPOD was first outlined by \citet{lumley1967structure,lumley1970stochastic}. Early implementations of SPOD include  \citet{glauser1987coherent,glauser1992application,delville1994characterization,arndt1997proper,picard2000pressure,citriniti2000reconstruction} and \citet{gordeyev2000coherent}.  For statistically stationary flows, \citet{towne2018spectral} have demonstrated that the SPOD combines the advantages of POD, namely optimality and orthogonality, and DMD, namely temporal monochromaticity. In this work, we demonstrate how these properties can be leveraged for different applications. 

In the past, SPOD has been used to analyze a number of  turbulent flows, including jets \citep{arndt1997proper, gordeyev2000coherent,gordeyev2002coherent, gamard2002application, gamard2004downstream, jung2004downstream,    iqbal2007coherent, tinney2008low, tinney2008low2, schmidt2018spectral, pickering2019lift, nekkanti2020modal}, the wake behind a disk \citep{johansson2002proper, johansson2006far, johansson2006far2, tutkun2008three, ghate2020broadband,nidhan2020spectral}, pipe flows \citep{hellstrom2014energetic,hellstrom2015evolution,hellstrom2016self, hellstrom2017structure}, and channel flows \citep{muralidhar2019spatio}. Several studies have shown that a significant amount of energy is captured by the first few modes at each frequency. For jets and disk wakes, \citet{glauser1987coherent}, \citet{citriniti2000reconstruction},  \citet{jung2004downstream}, \citet{johansson2006far2}, \citet{tinney2008low2} have shown that the leading mode and first three modes capture at least $40\%$, and $80\%$ of the total energy, respectively. Partial reconstructions of the flow field from SPOD were previously shown by \citet{citriniti2000reconstruction}, \citet{tinney2008low2}, and more recently, by \citet{ghate2020broadband}. We demonstrate low-rank reconstruction using two approaches. One is by inverting the SPOD algorithm, which was previously employed by \citet{citriniti2000reconstruction} in a similar manner, and for which we present an alternative means of computation based on convolution in the time domain. The other one is by taking an oblique projection of the data on the SPOD modes. The advantages and disadvantages of both approaches for different applications are discussed.

The first of these applications is denoising. Most experimental flow field data exhibit measurement noise that hampers the physical analysis. The computation of spatial derivatives required for quantities like the vorticity or strain rate, for example, leads to particularly large errors. Another difficulty is that physically relevant small-scale structures may be concealed by the noise. The most wide-spread experimental technique for multi-dimensional flow field measurement is particle image velocimetry (PIV). Common techniques to remove noise from PIV data include spatial filtering \citep{discetti2013spatial}, temporal filtering based on Fourier truncation, and POD-based techniques \citep{raiola2015piv, brindise2017proper}. Spatial filters are typically based on Gaussian smoothing \citep{discetti2013spatial}. Different temporal filters such as median filters \citep{son2001evaluation}, Hampel filters \citep{fore2005nonlinear}, Wiener filters \citep{vetel2011denoising}, and band-pass filters \citep{sciacchitano2014elimination} have been used for denoising PIV data. A comparison of different spatial and temporal filters is presented, for example, in \citet{vetel2011denoising}.  As an alternative to these standard techniques, POD reconstruction has been used as a means of denoising through mode truncation. Low-dimensional reconstructions from standard POD have been applied for this purpose to flows past a backward-facing step \citep{kostas2005comparison}, arterial flows \citep{charonko2010vitro,brindise2017hemodynamics}, turbulent wakes \citep{raiola2015piv}, and vortex rings \citep{stewart2012vortex, brindise2017proper}. In this contribution, we demonstrate the use of SPOD for denoising on surrogate data obtained by imposing high levels of additive Gaussian noise on simulation data. We demonstrate that SPOD-based denoising combines certain advantages of temporal filters and standard POD-based denoising.

Due to their chaotic nature, turbulent flows are characterized by high levels of intermittency. A common tool for the analysis of intermittent behaviour is frequency-time analysis, that is, the representation of the frequency content of a time signal as a function of time. This representation is particularly well-suited to identify events such as short-time interval of high or low energy, and identify their wave characteristics, i.e., frequencies or wavenumbers. Frequency-time analysis can be performed using several different signal-processing tools such as wavelet transforms (WT) \citep{farge1992wavelet}, the short-time Fourier transform (STFT) \citep{cohen1995time}, the S-transform \citep{stockwell1996localization}, the Hilbert-Huang transform \citep{huang1998empirical}, and the Wigner-Ville distribution \citep{boashash1988note}. Short-time Fourier and wavelet transforms are arguably the most widely used techniques in fluid mechanics. Frequency-time diagrams obtained from these methods are generally referred to as spectrograms, or as scalograms for wavelet transforms. STFT performs Fourier transforms on consecutive short segments of a time signal. It has been used, for example, in the analysis of blood flows \citep{izatt1997vivo,zhang2003comparison}, magnetohydrodynamics \citep{bale2005measurement}, aerodynamics \citep{samimy2007feedback}, and physical oceanography \citep{brown1989observations}.  The wavelet transform is based on the convolution of the time signal with a compact waveform, the so-called mother wavelet, that is scaled to represent different frequencies. Typical applications of the WT are found in atmospheric science \citep{gu1995secular}, oceanography \citep{meyers1993introduction}, and, most importantly for the present work, turbulence research \citep{farge1992wavelet}.  In the latter context, they have been used to extract coherent structures \citep{farge1999non,farge2001coherent}, and analyze their intermittency \citep{camussi1997orthonormal, onorato2000small, camussi2002coherent}.  All methods mentioned above are signal-processing techniques that are applied to one-dimensional data. Here, we expand on the ideas of \citet{schmidt2017wavepacket} and \citet{towne2019time}, and analyse the intermittency of the entire flow field. The underlying idea is that the global dynamics of the entire flow field can be described in terms of a limited set of statistically prevalent, most energetic coherent flow structures. For statistically stationary flows, such structures are distilled by the SPOD, and their temporal dynamics are described by the SPOD expansion coefficients. Since SPOD is a frequency-domain technique, this idea leverages the fact that each SPOD mode is associated with a single frequency. Based on the two reconstruction techniques mentioned above, we apply analyze and compare two variants of SPOD-based frequency-time analysis. The frequency-domain approach relies on direct inversion of the SPOD algorithm and was previously demonstrated by \citet{towne2019time}. This approach, however, becomes computationally intractable even for moderate-sized, two-dimensional data. We show that this problem can be avoided by the convolution-based approach introduced herein. 

Prewhitening is post-processing technique for trend detection that is commonly used in the atmospheric and geophysical sciences. It was first proposed by \citet{von1995misuses}. Prewhitening was used, for example, for the detection of trends in temperature and precipitation data \citep{zhang2001trends}, rainfall \citep{lacombe2012drying}, teleconnections \citep{rodionov2006use}, and hydrological flows \citep{khaliq2009identification, serinaldi2015stationarity}. Technically, prewhitening is achieved by a filtering operation that results in a flat power spectrum to remove serial correlations. We show two different SPOD-based ways to achieve this goal in the frequency domain.

The remainder of this paper is organized as follows. \autoref{sec:method} describes the two techniques for SPOD-based flow field reconstruction. In \autoref{sec:results}, we demonstrate these techniques on numerical data of a turbulent jet. The four different applications, SPOD-based low-dimensional reconstruction, denoising, frequency-time analysis, and prewhitenening are demonstrated in \autoref{low-dimensional reconstruction}, \autoref{denoising}, \autoref{frequency-time analysis}, and \autoref{appendix prewhitening}, respectively. \autoref{sec:conclusion} summarizes this work.

\section{Methodology } \label{sec:method}

\subsection{Spectral proper orthogonal decomposition (SPOD)}
In the following, we provide an outline of a specific procedure of computing the SPOD based on Welch's method \citep{welch1967use} and emphasize aspects that are important in the context of data reconstruction. Refer to the work of \citet{towne2018spectral} for details of the derivation and mathematical properties and \citet{schmidt2020guide} for a practical introduction to the method.  

Given a fluctuating flow field $\vb{q}_{i}=\vb{q}(t_i)$, where $i=1,\cdots,{n_t}$, that is obtained by subtracting the temporal mean $\Bar{\vb{q}}$ from each snapshot of the data, we start by constructing a snapshot matrix
\begin{equation}
\vb{Q}= \bqty{ \vb{q}_1, \vb{q}_2, \cdots , \vb{q}_{n_t} }.
\label{eq 1}
\end{equation}
Note that multi-dimensional data is cast into the form of a vector $\vb{q}_{i}$ of length $n$, corresponding to the number of variables times the number of grid points. The instantaneous energy of each time instant, or snapshot, is expressed in terms of a spatial inner product 
\begin{equation}
\|\vb{q}\|_x^2=\big<\textbf{q},\textbf{q} \big>_x = \int_{\Omega} \textbf{q}^*(x',t)\textbf{W}(x')\textbf{q}(x',t) dx',
\label{eq 2}
\end{equation}
where $\textbf{W}$ is a positive-definite Hermitian matrix that accounts for the component-wise weights, $\Omega$ the spatial domain of interest, and $(.)^*$ denotes the complex conjugate. The common form of space-only POD is obtained as the eigendecomposition of $\vb{Q}\vb{Q}^*\vb{W}$ and yields modes that are optimal in terms of equation (\ref{eq 2}). SPOD, however, specializes POD to statistically stationary processes and seeks modes that are optimal in terms of the space-time inner product 
\begin{equation}
\|\vb{q}\|_{x,t}^2=\big<\textbf{q},\textbf{q} \big>_{x,t} = \int_{-\infty}^{\infty} \int_{\Omega} \textbf{q}^*(x',t)\textbf{W}(x')\textbf{q}(x',t) \dd x' \dd t.
\label{eq 3}
\end{equation}
For statistically stationary data, it is natural to proceed in the frequency domain and solve the POD eigenvalue problem for Fourier transformed two-point space-time correlation matrix, that is, the cross-spectral density matrix. To estimate the CSD, the data is segmented into $n_{\rm blk}$ overlapping blocks with $n_{\rm fft}$ snapshots in each of them.
\begin{equation}
{\vb{Q}}^{(k)}=\bqty{\vb{q}_1^{(k)}, \vb{q}_2^{(k)}, \cdots, \vb{q}_{n_{\rm fft}}^{(k)} } 
\label{eq 4}
\end{equation}
If the blocks overlap by $n_{\rm ovlp}$ snapshots, the $j$-th column in the $k$-th block is given by 
\begin{equation}
\vb{q}_{j}^{(k)}=\vb{q}_{j+(k-1)(n_{\rm fft}-n_{\rm ovlp})+1}.
\label{eq 5}
\end{equation}
Each block is considered as a statistically independent realization of the flow under the ergodic hypothesis. The motive behind the segmentation of data is to increase the number of ensemble members. In practice, a windowing function is applied to each block to reduce spectral leakage. In this study, we use the symmetric hamming window
\begin{equation}
w(i+1)=0.54-0.46 \cos{\pqty{\frac{2 \pi i}{n_{\rm fft} -1} }} \quad\text{for}\quad i=0,1, \cdots, n_{\rm fft}-1.
\label{eq 6}
\end{equation}
Following best practices established by \citet{harris1978use}, we only apply  windowing for overlapping blocks to avoid excessive loss of information at the boundaries, i.e, if $n_{\rm ovlp} \neq 0$. Subsequently, the weighted temporal discrete Fourier transform,
\begin{equation}
\Hat{\vb{q}}_{j}^{(k)}=\mathcal{F}\qty{w(j)\vb{q}_{j}^{(k)}},
\label{eq fft}
\end{equation}
is performed on each windowed block to obtain the Fourier-transformed data matrix,
\begin{equation}
\Hat{{\vb{Q}}}^{(k)}=\bqty{\Hat{\vb{q}}_1^{(k)}, \Hat{\vb{q}}_2^{(k)}, ..., \Hat{\vb{q}}_{n_{\rm fft}}^{(k)} }  ,
\label{eq 7}
\end{equation}
where $\Hat{\vb{q}}_i^{(k)}$ denotes the $k$-th Fourier realization at the $i$-th discrete frequency. Next, we reorganize the data by frequency. The matrix containing all realizations of the Fourier transform at the $l$-th frequency reads
\begin{equation}
\Hat{\vb{Q}}_{l}=\bqty{\Hat{\vb{q}}_{l}^{(1)}, \Hat{\vb{q}}_{l}^{(2)}, ... \Hat{\vb{q}}_{l}^{(n_{\rm blk})} }.
\label{eq 8}
\end{equation}
From this form, the SPOD modes, $\vb*{\Phi}$,  and associated energies, $\lambda$, can be computed as the eigenvectors and eigenvalues of the CSD matrix $\vb{S}_l= \Hat{\vb{Q}}_{l}  \Hat{\vb{Q}}_{l}^{*}$. In practice, the number of spatial degrees of freedom, $n$, is often much larger than number of realizations. In that case, it is more economical to solve the analogous eigenvalue problem

\begin{equation}
\frac{1}{n_{\rm blk}}\vb{\Hat{Q}}_{l}^{*}\vb{W} \vb{\Hat{Q}}_{l} \vb*{\Psi}_{l}=\vb*{\Psi}_{l} \vb*{\Lambda}_{l}
\label{eq 11}
\end{equation}
for the coefficients $\vb{\psi}$ that expand the SPOD modes in terms of the Fourier realizations. In terms of the column matrix $\vb*{\Psi}_{l}=[\psi_l^{(1)}, \psi_l^{(2)}, \cdots \psi_l^{(n_{\rm blk})}]$, the SPOD modes at the $l$-th frequency are recovered as

\begin{equation}
\vb*{\Phi}_{l} = \frac{1}{\sqrt{n_\textrm{blk}}}\vb{\Hat{Q}}_{l} \vb*{\Psi}_{l} \vb*{\Lambda}_{l}^{-1/2}.
\label{eq 12}
\end{equation}
The matrices $\vb*{\Lambda}_{l}=\text{diag} ( \lambda_{l}^{(1)}, \lambda_{l}^{(2)}, \cdots , \lambda_{l}^{(n_{\rm blk})} ) $, where by convention $ \lambda_{l}^{(1)} \ge \lambda_{l}^{(2)} \ge \cdots  \ge \lambda_{l}^{(n_{\rm blk})} $, and $\vb*{\Phi}_{l}=[ \vb*{\phi}_{l}^{(1)}, \vb*{\phi}_{l}^{(2)}, \cdots , \vb*{\phi}_{l}^{(n_{\rm blk})} ] $ contain the SPOD energies and modes, respectively. By construction, the SPOD modes are orthogonal in the space-time inner product, equation (\ref{eq 3}). At any given frequency, the modes are also orthogonal in the spatial inner product, equation (\ref{eq 2}).

\subsection{Data reconstruction} \label{sec:reconst}

In  \S \ref{sec:reconst freq}, we show how the data can be reconstructed in the frequency-domain, that is, the inversion of the SPOD. An alternative approach based on (oblique) projection in the time domain is presented in \S\ref{sec:reconst time}. Whether one approach or the other is preferred depends on the specific application. Detailed discussions for each application under consideration in this work can be found in \S\ref{sec:results}. We will use the term \textit{frequency-domain} if the expansion coefficients are computed using the inversion of the SPOD problem, equation (\ref{eq 15}), and \textit{time-domain} if oblique projection, equation (\ref{eq 22}), is used.

\subsubsection{Reconstruction in the Frequency Domain}\label{sec:reconst freq}
The common numerator of the different applications of the SPOD considered in this study is that they require truncation or re-weighting of the SPOD basis. In practice, this is achieved by modifying the expansion coefficients. The original realizations of the Fourier transform at each frequency can be reconstructed as
\begin{equation}
\Hat{\vb{Q}}_{l}= \vb{\boldsymbol{\Phi}}_l \vb{A}_l,
\label{eq 14}
\end{equation}
where $\vb{A}_l$ is the matrix of expansion coefficients,
\begin{equation}
\vb{A}_{l} =\sqrt{n_{\rm blk}} \vb*{\Lambda}^{1/2}_{l} \vb*{\Psi}^*_{l}  = \vb{\boldsymbol{\Phi}}_l^{*} \vb{W} \Hat{\vb{Q}}_{l}.
\label{eq 15}
\end{equation}
Equation (\ref{eq 15}) shows that the expansion coefficients can either be saved during the computation of the SPOD or be recovered later by projecting the Fourier realizations onto the modes. In the following, we omit the frequency index $l$ with the understanding that the SPOD eigenvalue problem is solved at each frequency separately. From equation (\ref{eq 14}), it can be inferred that each column of the matrix
\begin{equation}
      \begin{aligned}
     \vb{A} = \left[\begin{array}{cccc}   
    {a}_{11} & {a}_{12} &\cdots & {a}_{1 n_{\rm blk}} \\
        {a}_{21} & {a}_{22} &\cdots & {a}_{2 n_{\rm blk}} \\
     \vdots & \vdots &\ddots &\vdots  \\  
    {a}_{n_{\rm blk} 1} & {a}_{n_{\rm blk} 2} &\cdots & {a}_{n_{\rm blk} n_{\rm blk}}
  \end{array}\right],
      \end{aligned}
      \label{eq 17}
\end{equation}
contains the expansion coefficients that allow for the reconstruction of a specific Fourier realization from the SPOD modes. \emph{Vice versa}, the coefficients contained in each row of $\vb{A}$ can be used to expand a specific SPOD mode in terms of the Fourier realizations. This can most easily be seen by rewriting equation (\ref{eq 14}) as  $\vb{\boldsymbol{\Phi}}_l = \frac{1}{n_{\rm blk}}\Hat{\vb{Q}}_{l}\vb{A}_l^*\vb*{\Lambda}^{-1}_l$. 
The Fourier-transformed data of the $k$-th block can be reconstructed as
\begin{equation}
\Hat{\vb{Q}}^{(k)}= \bqty{ \pqty{\sum\limits_{i} a_{ik} \vb*{\phi}^{(i)}}_{\!\!\!l=1} , \pqty{\sum\limits_{i} a_{ik} \vb*{\phi}^{(i)}}_{\!\!\!l=2} , \cdots, \pqty{\sum\limits_{i} a_{i k} \vb*{\phi}^{(i)}}_{\!\!\!l=n_{\rm fft}}} .
\label{eq 19}
\end{equation}

The original data in the $k$-th blocks ${\vb{Q}}^{(k)}$ can now be recovered using the inverse (weighted) Fourier transform,
\begin{equation}
{\vb{q}}_{j}^{(k)}=\frac{1}{w(j)}\mathcal{F}^{-1}\qty{\Hat{\vb{q}}_{j}^{(k)}}.
\label{eq ifft}
\end{equation}
Reconstructing the time series from the reconstructed data segments concludes the inversion of the SPOD.

\begin{figure}
\centering
{\includegraphics[trim={0cm 2.4cm 0cm 3.5cm },clip,width=1.0\textwidth]{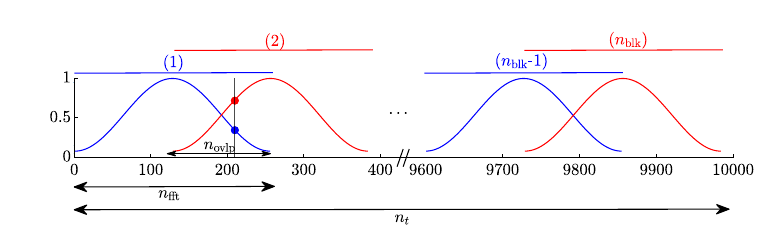}}
\caption{Schematic representation of overlapping blocks, hamming window and the SPOD parameters such as $n_t$, $n_{\rm{fft}}$, $n_{\rm{ovlp}}$ and  $n_{\rm{blk}}$. The left, and right blocks are denoted using blue, and red lines, respectively. The vertical line corresponds to the snapshot number 210, and the blue and red circles indicate the corresponding value of the window function.} 
\label{fig windowing function}
\end{figure}

A schematic of the windowing and blocking strategy is shown in figure \ref{fig windowing function}. The use of overlapping blocks leads to an ambiguity in the reconstruction as the $i$-th snapshot can either be obtained from the $k$-th block as ${\vb{q}}^{(k)}_{j}$, or from the  $(k+1)$-th block as ${\vb{q}}^{(k+1)}_{j+n_{\rm ovlp}- n_{\rm fft}}$. Different possibilities to remove this ambiguity are described in appendix \ref{appendix 1}. Based on this discussion, snapshots are reconstructed as averages of two reconstructions from overlapping blocks, weighted by the relative value of their windowing function. Partial reconstructions in the frequency-domain are readily achieved by zeroing the expansion coefficients of specific modes prior to applying the inverse Fourier transform.

\subsubsection{Reconstruction in the Time Domain}\label{sec:reconst time}
As an alternative to the reconstruction in the frequency domain, we present in the following a projection-based approach in the time domain. This approach is computationally efficient, and has the advantage that it can be applied to new data that was not used to compute the SPOD and to individual snapshots. However, the time-domain reconstruction does not leverage the orthogonality of the SPOD modes in the space-time inner product. Instead, it is based on an oblique projection of the data onto the modal basis. We start by representing the data as a linear combination of the SPOD modes as
\begin{equation}
\vb{Q}\approx   \tilde{\vb{\boldsymbol{\Phi}}}  \tilde{\vb{A}}.
\label{eq 20}
\end{equation}
The matrix $\tilde{\vb{\boldsymbol{\Phi}}}$ contains the basis of SPOD modes at all frequencies. Arranging the basis vectors by frequency first, we write 
\begin{equation}
\tilde{\vb*{\Phi}}= \bqty{\vb{\boldsymbol{\phi}}_{1}^{(1)}, \vb{\boldsymbol{\phi}}_{1}^{(2)}, \cdots , \vb{\boldsymbol{\phi}}_{1}^{(n_{\rm blk})}, \vb{\boldsymbol{\phi}}_{2}^{(1)}, \vb{\boldsymbol{\phi}}_{2}^{(2)}, \cdots , \vb{\boldsymbol{\phi}}_{2}^{(n_{\rm blk})}, \cdots, \vb{\boldsymbol{\phi}}_{n_{\rm fft}}^{(1)}, \vb{\boldsymbol{\phi}}_{n_{\rm fft}}^{(2)}, \cdots , \vb{\boldsymbol{\phi}}_{n_{\rm fft}}^{(n_{\rm blk})} }.
\label{eq 21}
\end{equation}
Assuming that $\tilde{\vb*{\Phi}}$ has full column rank, the matrix of expansion coefficient is obtained from the weighted oblique projection
\begin{equation}
\tilde{\vb{A}} = \pqty{\tilde{\vb{\boldsymbol{\Phi}}}^{*} \vb{W} \tilde{\vb{\boldsymbol{\Phi}}}}^{-1}  \tilde{\vb{\boldsymbol{\Phi}}}^{*} \vb{W} \vb{Q}. 
\label{eq 22}
\end{equation}
The oblique projection is required as SPOD modes at different frequencies are not orthogonal in the purely spatial inner product, $\langle\cdot,\cdot\rangle_x$, defined in equation (\ref{eq 2}). We furthermore use a weighted oblique projection based on the weight matrix $\vb{W}$ to guarantee compatibility with this inner product. Using the oblique projection, a single snapshot $\vb{q}=\vb{q}(\vb{x},t)$ is represented in the SPOD basis as
\begin{equation}
\tilde{\vb{q}} = \tilde{\vb{\boldsymbol{\Phi}}} \pqty{\tilde{\vb{\boldsymbol{\Phi}}}^{*} \vb{W} \tilde{\vb{\boldsymbol{\Phi}}}}^{-1}  \tilde{\vb{\boldsymbol{\Phi}}}^{*} \vb{W} \vb{q}, 
\label{eq 23}
\end{equation}
and the entire data is recovered as 
\begin{equation}
\tilde{\vb{Q}} = \tilde{\vb{\boldsymbol{\Phi}}} \tilde{\vb{A}}.
\label{eq 24}
\end{equation}
In the case where $\tilde{\vb*{\Phi}}$ is rank deficient or ill-conditioned, $\tilde{\vb*{\Phi}}^{*} \vb{W} \tilde{\vb{\boldsymbol{\Phi}}}$ is not invertible. For a generally rank-$r$ matrix, we perform the symmetric eigenvalue decomposition, 
\begin{equation}
    \tilde{\vb*{\Phi}}^{*} \vb{W} \tilde{\vb{\boldsymbol{\Phi}}} = \vb{U} \vb{D} \vb{U}^{*},
\quad\text{with}\quad \vb{U}=\left[
\vb{U}_1 \; \vb{U}_2\right],\; \vb{D}=\mqty[
\vb{D}_1 & \mathbf{0} \\
\mathbf{0} & \vb{0}], \label{eq 25}
\end{equation}
where $\vb{D}_1=\textrm{diag}(d_1,d_2, \cdots, d_r)$ with $d_1\geq d_2\geq\dots\geq d_r>0$ is the diagonal matrix of (numerically) non-zero eigenvalues and $\vb{U}_1$ is the corresponding matrix of orthonormal eigenvectors. In some applications, we desire a more aggressive truncation to rank $k<r$. For full-rank reconstructions, we use a truncation threshold of $\frac{d_k}{d_1} = 10^{-6}$ in this work. Denoted by
\begin{equation}
\tilde{\vb{A}}_{\{k\}} =  \vb{U}_{\{k\}} \vb{D}_{\{k\}}^{-1} \vb{U}_{\{k\}}^{*} \tilde{\vb*{\Phi}}^{*} \vb{W} \vb{Q}
\label{eq 23}
\end{equation}
is the rank-$k$ approximation of $\tilde{\vb{A}}$, where $\vb{D}_{\{k\}}=\textrm{diag}(d_1,d_2, \cdots, d_k)$ and $\vb{U}_{\{k\}}=\qty[\vb{u}_1,\vb{u}_2, \cdots, \vb{u}_k]$. In the truncated basis, $ \vb{U}_{\{k\}} \vb{D}_{\{k\}}^{-1} \vb{U}_{\{k\}}^{*}$ approximates $\pqty{\tilde{\vb*{\Phi}}^{*} \vb{W} \tilde{\vb{\boldsymbol{\Phi}}}}^{-1}$.
In the following, we demonstrate how the above and other truncation and partial reconstruction strategies can be used to achieve a number of objectives in the processing of flow data.

\section{Applications of SPOD: Low-rank reconstruction, denoising, frequency-time analysis, and prewhitening} \label{sec:results}
In this section, four different uses of SPOD-based applications are introduced and demonstrated on the example of a turbulent jet. The theoretical background of each applications is presented in the context of the jet. In particular, low-dimensional reconstruction is discussed in \S \ref{low-dimensional reconstruction}, denoising in \S \ref{denoising}, frequency-time analysis in \S \ref{frequency-time analysis}, and prewhitening in \S \ref{appendix prewhitening} respectively.


We consider the LES data of an isothermal subsonic turbulent jet, the Reynolds number, Mach number and temperature ratio are defined as $Re=\rho_j U_j D/\mu_j =0.45 \times 10^6$, $M_j= U_j/c_j=0.4$ and $T_j/T_{\infty}=1.0$, respectively, where $\rho$ is the density, $U$ velocity, $D$ nozzle diameter, $\mu$ dynamic viscosity, $c$ speed of sound, and $T$ temperature. The subscripts $j$ and $\infty$ refer to the jet inlet and free-stream conditions, respectively. We use the large-eddy simulation data computed by \citet{bres2019modelling}.  The simulation was performed using the compressible flow solver `Charles' \citep{bres2017unstructured} on an unstructured grid using a finite-volume method. The reader is referred to Br{\`e}s et al. (\citeyear{bres2018importance,bres2019modelling}) for further details on the numerical method. The LES database consists of 10,000 snapshots sampled at an interval of $\Delta t c_{\infty}/D=0.2$ acoustic time units. Data interpolated on a cylindrical grid spanning $x,r$ $\in$ $[0, 30] \times [0, 6]$ is used in this analysis. The flow is non-dimensionalized by the nozzle exit values, namely, velocity by $U_j$, pressure by $\rho_j U_j^2$, length by the nozzle diameter $D$, and time by $D/U_j$. Frequencies are reported in terms of the Strouhal number $\St=f D/U_j$. For simplicity, and without loss of generality, we perform our analysis only on the pressure field in what follows. Refer to \citet{freund2009turbulence} for analysis on different energy norms and POD modes of a jet. We further exploit the rotational symmetry of the jet and consider individual azimuthal Fourier components. The helical ($m=1$) component provides an example of complex data.

\begin{figure}
\centering
{\includegraphics[trim={0cm 0.5cm 0cm 0.5cm },clip,width=1.0\textwidth]{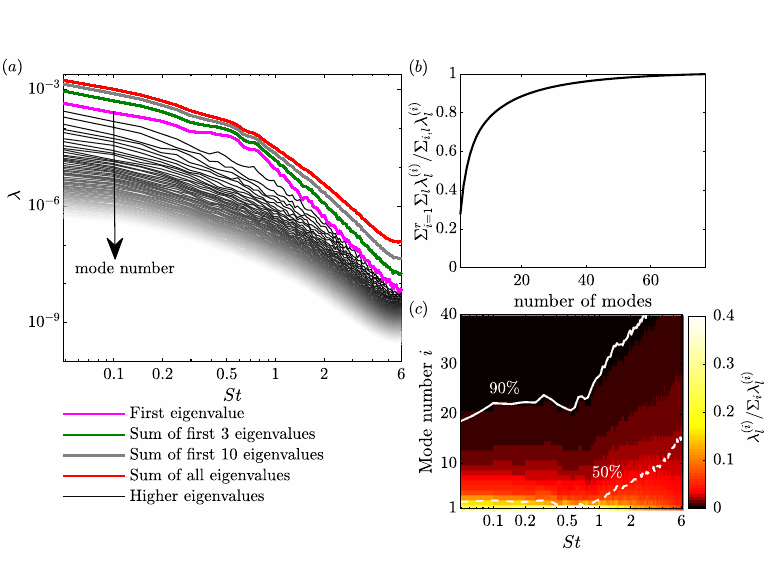}}
\caption{SPOD spectra of the turbulent jet for $m=1$. All eigenvalues (gray lines) and the sum of all eigenvalues (red line), corresponding to the integral PSD, are shown. The normalized cumulative energy content and the percentage of energy accounted by each mode as a function of frequency are shown in ($b$) and ($c$), respectively. The solid and dashed white lines indicate the number of modes required to retain $90\%$ and $50\%$ energy at each frequency.}
\label{fig 1}
\end{figure}

To determine the spectral estimation parameters for the SPOD, we follow the guidelines provided in \citet{schmidt2020guide} and \citet{schmidt2018spectral}, which, in turn, follow standard practice in spectral estimation. The SPOD is computed for blocks containing $n_{\rm fft} = 256 $ snapshots with $50\%$ overlap, resulting in a total number of  $n_{\rm blk} =77$ blocks. A $50\%$ overlap is used to minimize the variance of the spectral estimate \citep{welch1967use}. 

Since SPOD yields one set of eigenpairs per frequency, we may investigate the contributions of different frequencies independently. The SPOD eigenvalues are represented in the form of a spectrum, reminiscent to a power spectrum, in figure \ref{fig 1}($a$). Gray lines of decreasing intensity connect eigenvalues of constant mode number and decreasing mode energy. The first, most energetic mode is shown in magenta. The red line represents the sum of all eigenvalues and corresponds to the power spectral density (PSD) integrated over the physical domain. This line of the integral total energy can be compared to truncated sums of eigenvalues, that is, the energy contained in reconstructions of different ranks. As the eigenvalues are sorted by energy, the lines corresponding to the truncated sums of the leading 3 and 10 eigenvalues fall between the leading-eigenvalue spectrum and the total energy curve.
Figure \ref{fig 1}($b$) shows the normalized cumulative energy content, independent of frequency. The first and the leading 10 modes contain $ 30 \%$ and $ 80 \%$ of the total energy, respectively. Figure \ref{fig 1}($c$) shows the percentage of energy accounted for by each mode as a function of frequency. At low frequencies, the first few modes contain a high percentage of energy, whereas the energy is more dispersed at higher frequencies. The solid and dashed white lines indicate the number of modes required to retain $50\%$ and $90\%$ of the total energy at each frequency.


\begin{figure}
\centering
{\includegraphics[trim={0cm 3.9cm 0cm 1cm },clip,width=1.0\textwidth]{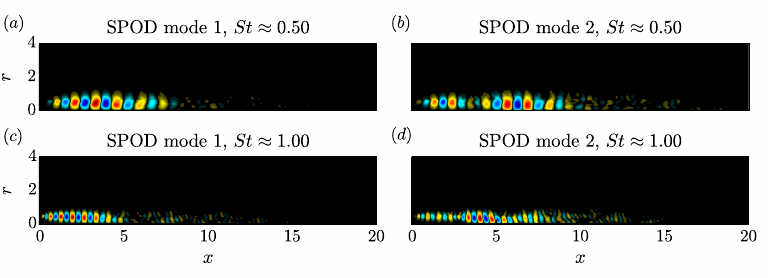}}
\caption{SPOD modes at $St=0.5$ ($a$,$b$) and at $St=1.0$ ($c$,$d$). The leading modes are shown in ($a$,$c$) and the suboptimal modes are shown in ($b$,$d$).}
\label{fig 2}
\end{figure}

The first and second modes at two representative frequencies are shown in figure \ref{fig 2}. The frequency $St=0.5$ corresponds to the maximum difference between first and second eigenvalues. The leading mode at this frequency shows a Kelvin-Helmholtz wavepacket \citep{suzuki2006instability, gudmundsson2011instability,schmidt2018spectral} in the shear layer of the jet.  A similar, but a more compact wavepacket structure is observed at $St=1.0$. The suboptimal mode at both frequencies exhibit a multi-lobed wavepacket structure, whose amplitude peaks in near the end of the potential core at $x\approx 6$. The reader is referred to \citep{schmidt2018spectral} and \citep{tissot2017wave} for a physical discussion of this observation and the link to non-modal instability. In the present context, our preliminary interest is in the desirable mathematical property of the SPOD that guarantee that the modes optimally represent the turbulent flow field in terms of the space-time inner product, equation (\ref{eq 3}). In the following, different uses of low-dimensional reconstructions that use SPOD modes as basis vectors are introduced and discussed.

\subsection{Low-dimensional flow field reconstruction} \label{low-dimensional reconstruction}
Since SPOD seeks an optimal series expansion for each frequency, the choice of what eigenpairs to include in a low-dimensional reconstruction is not obvious. Here we first discuss the most elementary way of truncation based on the frequency-wise optimality property, that is, a certain number of modes is retained at each frequency. We refer to this as a $n_\mathrm{modes} \times n_\mathrm{freq}$-mode reconstruction, where $n_\mathrm{modes}$ is the number of modes retained at each frequency, and $n_\mathrm{freq} = \frac{n_\mathrm{fft}}{2}+1$ is the number of positive frequencies, including zero. If all modes are linearly independent, the overall rank of the reconstructions is then given by the total number of basis vectors, $n_\mathrm{modes}n_\mathrm{freq}$. 

Following the discussion in \S \ref{sec:method}, we present two means of obtaining an SPOD-based low-dimensional reconstruction:
\begin{enumerate}
    \item in the frequency domain (see \S\ref{sec:reconst freq}) using equation (\ref{eq ifft}), and
    \item in the time domain (see \S\ref{sec:reconst time}) using equation (\ref{eq 23}). 
\end{enumerate}

The frequency-domain approach directly follows from the mathematical definition of the SPOD, and was previously used by
\citet{citriniti2000reconstruction,jung2004downstream,johansson2006far2, tinney2008low}. The time-domain approach can be viewed as the most general approach that can be applied to any given modal basis. It is not specific to SPOD, but commonly used for low-order modeling.   

\begin{figure} 
\centering
{\includegraphics[width=1.0\textwidth]{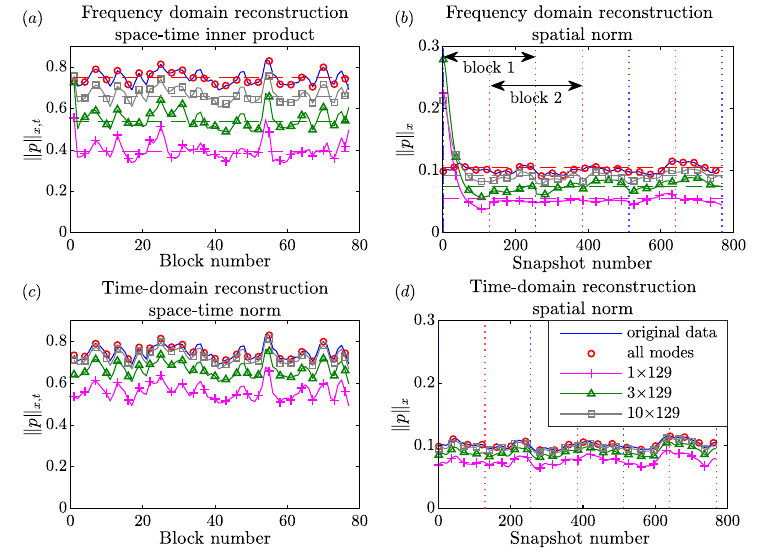}}
\caption{Low-dimensional reconstruction of the jet data: frequency-domain reconstruction ($a$,$b$), and time-domain reconstruction on the ($c$,$d$) in terms of the space-time norm ($a$,$c$) and the spatial norm ($b$,$d$). The original data (blue lines), shown for reference, is compared to the full reconstructions using all modes, and reconstructions using 10$\times$129, 3$\times$129, and 1$\times$129 modes. Summed SPOD mode energies are shown as dashed lines.  Vertical dotted blue and red lines in ($b$,$d$) indicate the left and right blocks in the reconstruction (see figure \ref{fig windowing function}).}
\label{fig 3}
\end{figure}

In the following, we first compare different low-dimensional reconstructions in terms of their block-wise and snapshot-wise energy in figure \ref{fig 3}. It follows from equation (\ref{eq 3}), that the energy of a single block is 
\begin{equation}
\|\textbf{q}\|_{x,t}^2 = \big<\textbf{q},\textbf{q} \big>_{x,t} = \int_{\Delta T} \int_{\Omega} \textbf{q}^*(x',t)\textbf{W}(x')\textbf{q}(x',t) \dd x' \dd t,
\label{eq 3.1}
\end{equation}
where $\Delta T = [t_{1+(k-1)(n_{\rm fft}-n_{\rm ovlp})},t_{{n_{\rm fft}}+(k-1)(n_{\rm fft}-n_{\rm ovlp})}]$ is the time interval of the $k$-th block. The spatial norm (\ref{eq 2}) measures the energy present in each snapshot. Both the block-wise and snapshot-wise pressure norms are computed for both reconstruction approaches. The evolution of the space-time norm is shown for the entire database in figure \ref{fig 3}($a$,$c$). For the spatial norm, we focus on the first 768 snapshots in figure \ref{fig 3}($b$,$d$). This segment corresponds to 5 blocks and exhibits dynamics that are representative of the rest of the data.

Low-dimensional reconstructions using 1$\times$129, 3$\times$129, 10$\times$129 modes and the reconstruction using all modes are presented in figure \ref{fig 3}. The dimension of the modal bases directly reflects their ability to capture the pressure norms of the data. The full-dimensional reconstructions in the frequency and time domain recover the data completely. Notably, the dynamics in space-time norm are accurately captured, even by the 1$\times$129-mode reconstruction. For a fixed number of modes, the time domain approach captures more energy and provides a better approximation of the data than the frequency domain approach. Take as an example the $10 \times 129$ basis: the time-domain reconstruction accurately approximates for the full data (figure \ref{fig 3}($c$,$d$)), which is notably under-predicted by frequency-domain reconstruction (figure \ref{fig 3}($a$,$b$)). This difference can be understood by considering the SPOD energy content of the reconstruction. The dashed lines in figure \ref{fig 3}($a$,$b$) denotes this energy, which is given by the sum of first $n_{\rm modes}$ eigenvalues over all frequencies. As expected, the space-time and spatial norm of the different reconstructions fluctuate about the sum of the eigenvalues. Higher energies are obtained by the time-domain reconstruction in figure \ref{fig 3}($c$,$d$) as the modal expansion coefficients obtained via oblique projection are not bound to specific frequencies. In what follows, we will see again and again that this flexibility of the expansion coefficients of the time-domain approach leads to an overall better reconstruction. This additional degree of freedom can be leveraged to obtain an accurate reconstruction of the flow dynamics. It is, in fact, the optimality property of the oblique projection, equation (\ref{eq 23}), that guarantees that the time-domain approach yields the best possible approximation in a least-square sense. For example, the 1$\times$129 time-domain reconstruction seen in figure \ref{fig 3}($d$) is sufficient to capture the dynamics of the original data accurately. From figure \ref{fig 3}($b$), it can be seen that the frequency-domain reconstruction significantly over-predicts the pressure norm during the first 32 snapshots. This effect only occurs for the two out-most blocks, which do not posses neighbouring blocks in one direction. In appendix \ref{appendix 1} (figure \ref{fig A2}), we show by comparison with rectangular window that this error is a result of the Hamming window. The presence of the windowing-effect in the first and last blocks is equally reflected in the space-time norm, see figure \ref{fig 3}($a$).

\begin{figure} 
\centering
{\includegraphics[trim={0cm 4.35cm 0cm 0cm },clip,width=1.0\textwidth]{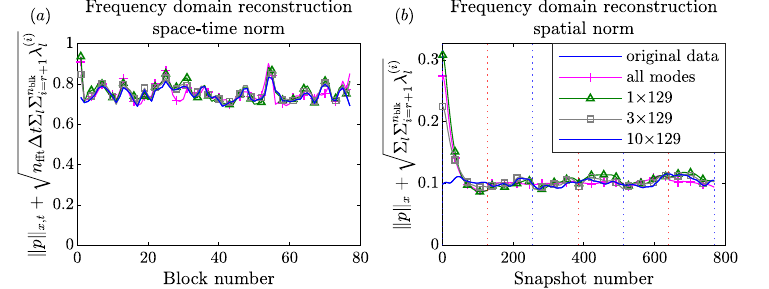}}
\caption{Low-dimensional reconstruction in the frequency-domain with truncation correction: ($a$) space-time norm; ($b$) spatial norm. Parts ($a$) and ($b$) correspond to figure \ref{fig 3}($a$) and ($b$), respectively, but with a correction for the truncated modes. The correction is facilitated by adding the energies of the truncated modes (given by their SPOD eigenvalues).}
\label{fig 4}
\end{figure}
Figure \ref{fig 4} demonstrates that the frequency-domain approach accurately recovers the mode energies given by the SPOD eigenvalues. By adding the residual energy contained in the truncated eigenvalues, both the space-time norm (figure \ref{fig 4}($a$)) and the spatial norm (figure \ref{fig 4}($b$)) can be collapsed to the total energy. The remaining differences are largely due to the windowing effect.

\begin{figure}
\centering
{\includegraphics[trim={0cm 1.05cm 0cm 0.25cm},clip,width=1.0\textwidth]{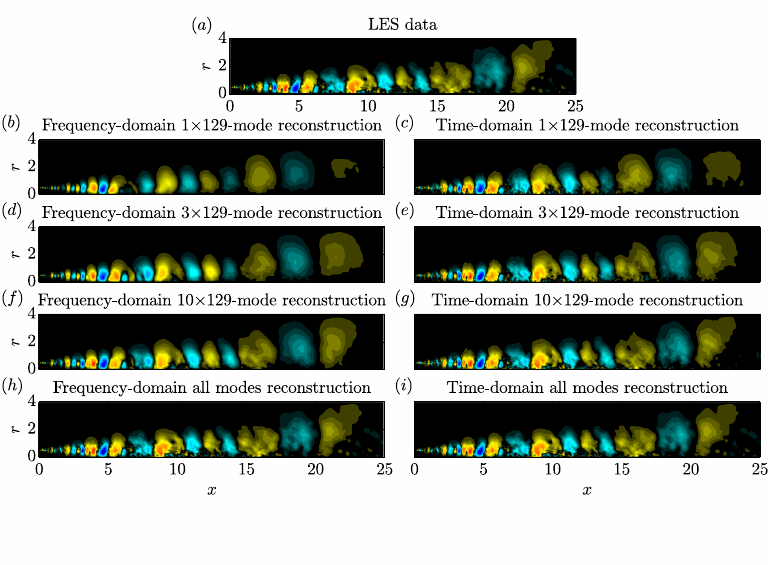}}
\caption{Instantaneous pressure field: ($a$) original flow field is shown; ($b$,$d$,$f$,$h$) reconstructions in the frequency domain; ($c$,$e$,$g$,$i$) reconstructions in the time domain. Flow fields reconstructed using 1$\times$129 modes, 3$\times$129 modes, 10$\times$129  modes, and all modes are shown in ($b$,$c$),  ($d$,$e$), ($f$,$g$), and ($h$,$i$), respectively.  Contours in ($a$-$i$) are reported on the same color axis.} 
\label{fig 5}
\end{figure}

Figure \ref{fig 5} compares a single time instant of the original data in ($a$) with reconstructions of increasing fidelity in the frequency (left) and the time domain (right). Both the 1$\times$129-mode reconstructions shown in figure \ref{fig 5}($b$,$c$) capture the dominant wavepackets. However, the frequency-domain reconstruction lacks the detail of the time-domain reconstruction. The higher accuracy of the time-domain reconstruction can be explained by its less stringent nature. As the leading SPOD modes often represent a spatially highly confined structure, other structures associated with the same frequency cannot be represented by the frequency-domain reconstruction. The KH-type wavepacket seen in figure \ref{fig 2}($c$) is a good example of such a confined structure. This difference between the approaches also explains the better reconstruction of the integral energy in the time domain, as previously observed in figure \ref{fig 3}. The higher-dimensional versions for both approaches shown in figure \ref{fig 5}($d$-$g$) yield increasingly more detailed and accurate reconstructions. Both approaches yield reconstructions that are indistinguishable from the original data when all modes are used (see figure \ref{fig 5}($h$,$i$)).


\begin{figure}
\centering
{\includegraphics[trim={0.4cm 4.5cm 0.5cm 0.0cm},clip,width=1.0\textwidth]{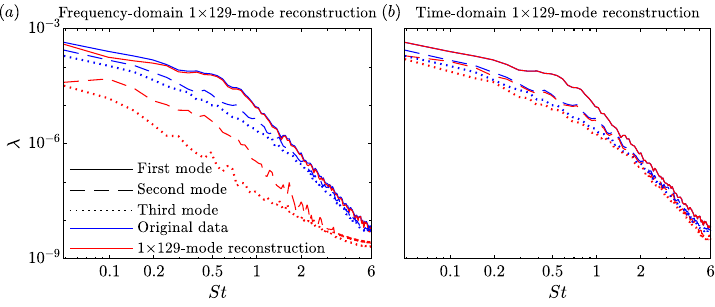}}\caption{
Comparison between SPOD eigenvalue spectra of the original data (blue lines) and 1$\times$129-mode reconstructions (red lines): ($a$) frequency-domain; ($b$) time-domain. Solid, dashed, and dotted lines denote the first, second, and third modes, respectively.} 
\label{fig 6}
\end{figure}
 
We infer from figures \ref{fig 3} to \ref{fig 5} that  reconstruction in the time-domain provides a better estimate of the flow field than the frequency-domain version. To understand this observation, figure \ref{fig 6} reports the SPOD eigenspectra of the 1$\times$129-mode reconstructions and compares them to those of the full data. Only the leading three eigenvalues are shown for clarity. The leading eigenvalue of the frequency-domain reconstruction in figure \ref{fig 6}($a$) approximately follows the full data with some discrepancies at lower frequencies. No such discrepancies are observed for the time-domain reconstruction in figure \ref{fig 6}($b$); in fact, the leading eigenvalue spectra are indistinguishable. Contrast this observation with the expectation that a $1\times n_\mathrm{freq}$ frequency-domain reconstruction should exactly reproduce the leading-mode eigenspectrum, and that all higher-eigenvalue spectra are expected to be zero. In the context of figure \ref{fig A3} in appendix \ref{appendix 1}, we show that this is an effect of windowing that is not observed when using a rectangular window. The time-domain reconstruction, on the other hand, accurately approximates the first, and, to some degree, the leading suboptimal eigenvalue spectra. This again demonstrates the higher accuracy of the time-domain reconstruction that results from the higher flexibility of the expansion.


 \begin{figure}
 \centering
{\includegraphics[trim={0.0cm 1.7cm 0.0cm 2.0cm},clip,width=1.0 \textwidth]{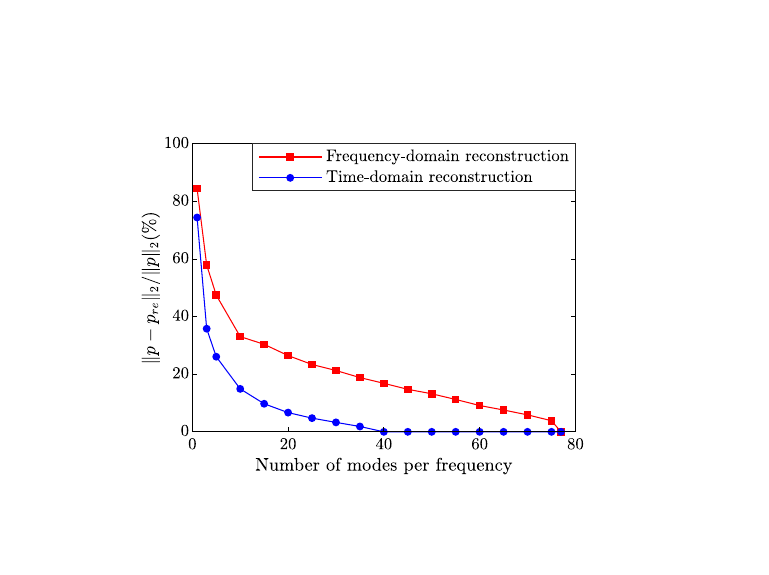}}
\caption{Two-norm error of the pressure field reconstructed with different rank approximation: (red lines with squares) reconstruction in the frequency domain; (blue lines with circles) reconstruction in the time domain.}
\label{fig 7}
\end{figure}

To quantify the accuracy of the two approaches, figure \ref{fig 7} compares their 2-norm errors as a function of the number of basis vectors (modes). For both the methods, the error reduces significantly as the number of modes retained per frequency increases from one to ten. For a fixed number of modes, the time-domain reconstruction is consistently more accurate. Recall that this is guaranteed by the optimality property of the oblique projection. It is, in fact, observed that the error of the time-domain reconstruction approaches machine precision for 40 or more modes per frequency. For the frequency-domain approach this only occurs for the full reconstruction using all modes.

\subsection{Denoising} \label{denoising}

After using SPOD truncation for low-dimensional approximation above, we now explore its potential for denoising. We will show that additive noise is captured by certain parts of the spectrum, and that truncation of these parts leads to efficient denoising. This strategy is most efficiently implemented in the frequency domain. The local-in-time optimality of the time-domain approach is a hindrance in this context as it tends to reconstruct the noise. On the contrary, the one-to-one correspondence between modes and frequencies of the frequency domain approach leads to efficient denoising.  

\begin{figure}
\centering
{\includegraphics[trim={0cm 2.2cm 0cm 2.25cm},clip, width=1.0 \textwidth]{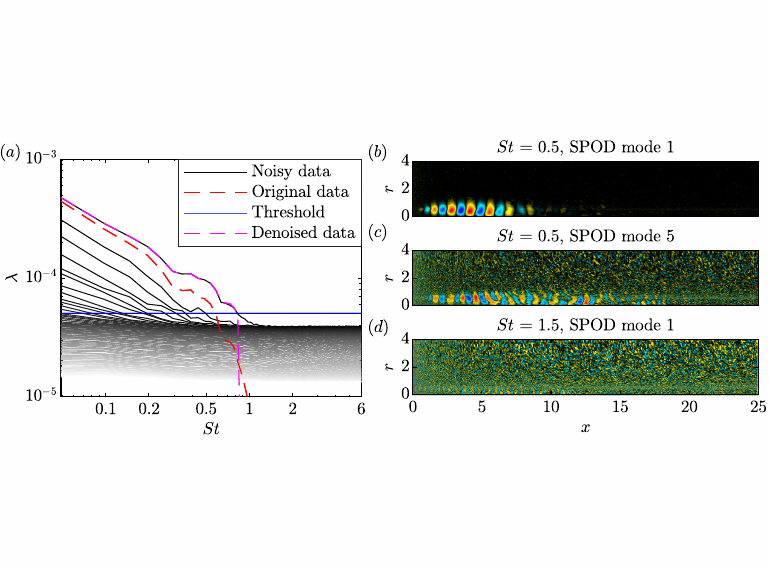}}
\caption{SPOD of data subjected to additive Gaussian white noise: ($a$) SPOD spectrum (black lines), leading SPOD eigenvalue of the original data (red dashed line), threshold of $5\times 10^{-5}$ (blue line), and the leading SPOD eigenvalue of the denoised data (magenta dashed line); ($b$) leading SPOD mode at $\St=0.5$; ($c$) fifth SPOD mode at $\St=0.5$; ($d$) leading SPOD mode at $\St=1.5$.  }
\label{fig 9}
\end{figure}

As the arguably most common type of noise occurring in experimental environments, we demonstrate denoising on additive Gaussian white noise. In particular, we add Gaussian white noise that has a standard deviation equal to the spatial mean of the standard deviation along the lipline of the pressure data. This scenario mimics, for example, heavily contaminated particle image velocimetry data in which the variance of the noise is of the same order as the variance of the physical phenomena of interest. The SPOD eigenvalue spectrum of this noisy data is shown in figure \ref{fig 9}($a$). Most noticeably, the addition of noise has introduced a noise floor at $\lambda\approx4\cdot 10^{-5}$, effectively cutting off the spectrum at $\St \approx 1.5$. Information above this frequency lies below the noise floor and is not directly accessible.  The leading eigenvalue of the original data (red dashed line) is shown for comparison. It lies well below the leading eigenvalue of the noisy data, which is elevated due to the energy contained in the added noise. To illustrate the ability of SPOD to differentiate between spatially correlated, physical structures and noise, examples of modes that are above and below the noise floor are compared in figure \ref{fig 9}($b$-$d$). The leading mode at $\St=0.5$ (figure \ref{fig 9}($b$)) clearly reveals the K-H wavepacket and is indistinguishable from the corresponding mode of the original data shown in figure \ref{fig 2}($a$) above. The fifth mode at $\St=0.5$ (figure \ref{fig 9}($c$)) and the leading mode at $\St=1.5$ (figure \ref{fig 9}($d$)), on the contrary, are heavily contaminated by noise. The noise floor in the SPOD spectrum is found to be a very good indicator of this distinction. We hence propose a denoising-strategy based on hard-thresholding of the spectrum. In this example, we pick a threshold of $5\times 10^{-5}$ (blue line), slightly above the noise floor. To address the effect of the truncation on the SPOD spectrum, we report the leading SPOD eigenvalue of the denoised data (magenta dashed line) in the same figure. It coincides with the leading eigenvalue of the noisy data up to the point where it intersects with the threshold limit, beyond which it falls off sharply, giving it the characteristics of a low-pass filter. A closer analysis of the truncated and original spectra reveals that the denoised field contains only $2.6\%$ energy of the noisy field, but that it contains $92.7\%$ energy of the original flow field. Another positive side effect is that only 74 out of 9933 modes ($n_{\rm freq} \times n_{\rm blk}$) have been retained, resulting in a space saving of $99.26\%$. These significant space savings are an advantage over standard denoising strategies based on low-pass filtering.

 \begin{figure}
\centering
{\includegraphics[trim={0cm 3.75cm 0cm 2.2cm },clip,width=1.0\textwidth]{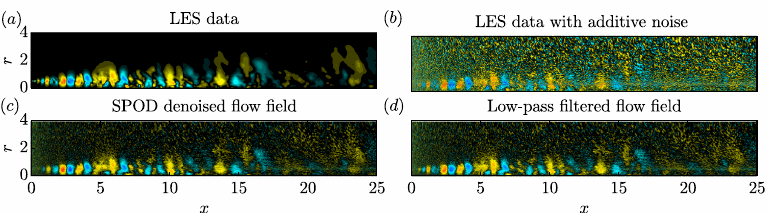}}
\caption{Comparison of noisy and denoised instantaneous pressure fields: ($a$) original; ($b$) data with additive Gaussian white noise; ($c$) SPOD-based denoised flow field; ($d$) low-pass filtered flow field. Denoising is achieved by rejecting all SPOD eigenpairs with $\lambda < 5\times 10^{-5}$. This hard threshold is indicated in figure \ref{fig 9}.}
\label{fig 10}
\end{figure}

A representative instantaneous snapshots of the original data is compared to its noisy and denoised counterparts, and to the result of standard low-pass filtering in figure \ref{fig 10}. The standard low-pass filter uses the cut-off frequency of $\St=0.8$ of the SPOD approach (inferred from figure \ref{fig 9}) in the truncation of the long-time Fourier transform. A higher threshold, and its associated cut-off frequency, was found to lead to more aggressive filtering that can partially remove relevant flow structures. A threshold below the noise floor, on the other hand, leads to unsatisfactory noise rejection. In practice, a good trade-off between noise rejection and preservation of physically relevant flow structures is achieved by using the SPOD spectrum as a gauge to choose a threshold slightly above the noise floor. A comparison of the denoised data in figure \ref{fig 10}($c$,$d$) with the noisy data in figure \ref{fig 10}($b$) shows that significant noise reduction was achieved in all parts of the domain using both strategies. The resulting denoised flow fields clearly reveal the flow structures present in the original data. By visual inspection of the filtered pressure fields shown in figures \ref{fig 10}($c$) and \ref{fig 10}($d$), the SPOD-based strategy appears somewhat more efficient at removing the noise.

\begin{figure}
\centering
{\includegraphics[trim={0cm 3cm 0cm 1.25cm },clip,width=0.95 \textwidth]{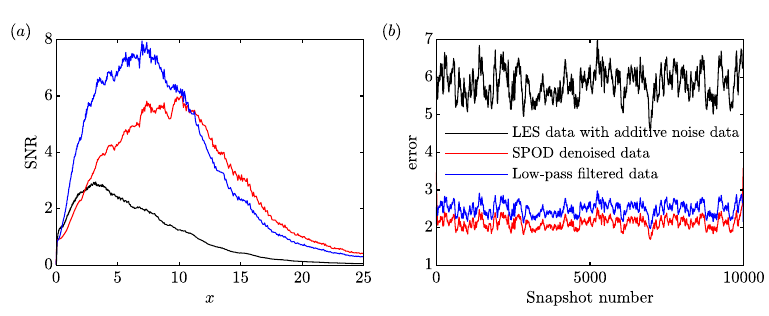}}
\caption{ Comparison of the two denoising strategies: ($a$) signal to noise ratio (SNR) along the lipline ($r=0.5$) for the noise added flow field (black line), SPOD-based denoised flow field (red lines), and the low-pass filtered flow field (blue line); ($b$) error of the noisy data, SPOD-based denoised data, and the low-pass filtered data. The amplitude of the additive noise was adjusted such that the average SNR along the lipline is one.}
\label{fig 11}
\end{figure}

For a more quantitative assessment, we compare the denoised flow fields in terms of two quantities. First, their signal-to-noise ratio (SNR) along the lipline, and second, the relative error between the denoised snapshots and the original data. The SNR is defined as
\begin{equation}
    {\rm SNR} = \frac{P_{\rm signal}}{P_{\rm noise}}= \frac{\sigma^2_{\rm signal}}{\sigma^2_{\rm noise}},
\label{eq 3.5}
\end{equation}
where $P$ is power and $\sigma$ standard deviation. We further define the integral (over the physical domain) error as \begin{equation}
    \text{error} =  \frac{\|\vb{q} - \vb{\check{q}}\|_x}{\|\vb{q}\|_x},
\end{equation}
where $\vb{q}$ and  $\vb{\check{q}}$ are the original and the denoised flow fields, respectively. Figure \ref{fig 11}($a$) compares the SNR along the lipline at $r=0.5$ for the noisy and the denoised flow fields. Both methods achieve to increase the signal-to-noise ratio over large parts of the domain. The low-pass filter performs better for $x\lesssim 10$, and the SPOD-based filter beyond that point. For $x\lesssim 2.5$, the SPOD-filtered pressure field exhibits a marginally lower SNR than the unfiltered data. We find that this is the results of the aggressive truncation of high-frequency components by the SPOD-based filter. A result that is almost identical to that of the low-pass filter can be achieved by lowering the $\lambda$-threshold (a similar value for both methods is used here for consistency). Figure \ref{fig 11}($b$) compares the time traces of the errors of the noisy and the two denoised flow fields. The error of the noisy data serves as a reference, and it is observed that both methods significantly reduce this error. The SPOD-based approach performs consistently  better than the low-pass filter. This result is consistent with the visual observation of the denoised fields in figure \ref{fig 10}($c$,$d$). Note that the threshold is an adjustable parameter in both methods. In practice, we find that by adjusting this parameter qualitatively very similar results can be obtained by both methods. This, however, leaves the SPOD-based approach with the advantage of significant data reduction.

\subsection{Frequency-time analysis} \label{frequency-time analysis} 

Intermittency, that is the occurrence of flow events at irregular intervals, is an inherent feature of any turbulent flow. A common approach for the characterization of intermittent behaviour is frequency-time analysis. The arguably most wide-spread tools of frequency-time analysis are wavelet transforms and short-time Fourier transform. Their outcomes are scalograms and spectrograms, respectively, that indicate the presence of certain scales (WT), or frequency components (STFT), at certain times. Both methods are signal-processing techniques that are applied to 1-D time series, and therefore only quantify intermittency locally. As an alternative to this local perspective, we demonstrate how SPOD expansion coefficients can be used to study the intermittency of the spatially coherent flow structures represented by the modes. Below, frequency-time analyses based on both,  time-domain and frequency-domain reconstructions, are introduced and compared. 


\subsubsection{Time-domain approach}
\begin{figure}
\centering
{\includegraphics[trim={0cm 3.45cm 0cm 1.75cm},clip, width=1.0\textwidth]{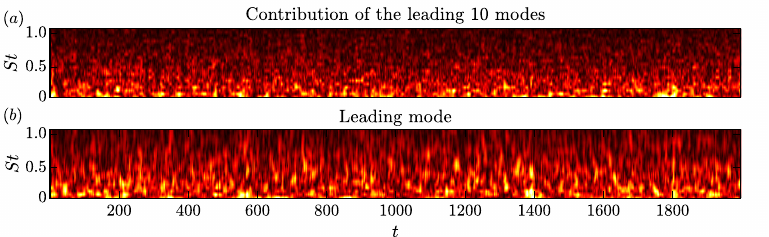}}
\caption{SPOD-based frequency-time diagrams obtained using the time-domain approach: ($a$) first 10 modes at each frequency; ($b$) leading mode at each frequency. The SPOD of the pressure field is considered.}
\label{fig 12}
\end{figure}

We first consider the time-domain approach, in which the expansion coefficients obtained via oblique projection readily describe the temporal behaviour of each mode. The amplitudes of the expansion coefficients computed from equation (\ref{eq 22}), $|\sum_{j=1}^{n_\text{modes}}\tilde{a}^{(j)}(f_l,t)|$, hence yields the desired frequency-time representation for the leading ${n_\text{modes}}$. The expansion coefficients are calculated using the full basis, i.e, $\tilde{\vb{\boldsymbol{\phi}}}$ in equation (\ref{eq 21}) consists of all modes at all frequencies. Subsequently, we only consider the expansion coefficients of the leading $n_{\rm{modes}}$ modes at each frequency.  An alternative approach is to perform the oblique projection using a reduced basis that consists of only the leading $n_{\rm{modes}}$ modes at each frequency. We find that the first approach is preferable in the context of frequency-time analysis and is explained in the appendix  \ref{appendix:frequnecy-time analysis } (see figure \ref{fig A5}).  
The frequency-time diagrams for $n_\textrm{modes}=10$, and 1, are shown in figure \ref{fig 12}($a$), and ($b$), respectively. The leading 10 modes correspond approximately to $80\%$ of the total energy as shown in  figure \ref{fig 1}($b$).  Most of the energy is concentrated at low frequencies, $St \lesssim 0.2$, as expected from the eigenvalue spectrum in figure \ref{fig 1}. The eigenvalue spectrum provides a statistical representation of the structures that are coherent in space and time, whereas the frequency-time diagrams provide a temporal information of these structures.  Bright yellow spots indicate high similarity of the instantaneous flow field with the leading mode in ($b$). These regions also correspond to high energy events as we will show in figure \ref{fig 12a}.

\begin{figure}
\centering
{\includegraphics[trim={0cm 3.35cm 0cm 1.75cm},clip, width=1.0\textwidth]{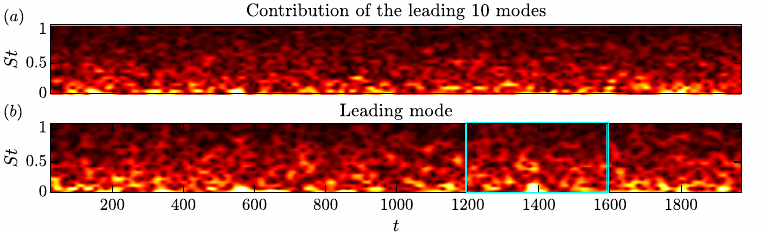}}
\caption{SPOD-based frequency-time diagrams obtained using the convolution approach: ($a$) first 10 modes at each frequency; ($b$) leading mode at each frequency. The SPOD of the pressure field is considered. The cyan box in ($b$) is centered around the global maximum of the instantaneous energy at $t=1380$, analysed in figures \ref{fig 12a} and \ref{fig 14}($c$,$d$) below.}
\label{fig 13}
\end{figure}

\subsubsection{Convolution-based approach}
A direct way of using SPOD for frequency-time analyses is in terms of the SPOD expansion coefficients. Since each block is associated with a finite time interval, this approach requires the computation of the SPOD using an overlap of $n_{\rm ovlp}= n_{\rm fft} - 1$ to obtain time-resolved coefficients \citep{towne2019time}. This approach assumes that the value of the expansion coefficient obtained from a finite time segment (block) represents the instantaneous frequency content at the center of the time segment. A limitation of this approach is its high memory requirement (3.6 TB for the present example). As an alternative, we propose a computationally tractable way of calculating time-continuous expansion coefficients based on the convolution theorem. Applying the convolution theorem to the inverse SPOD problem yields
    \begin{equation}
     \vb{a}_l^{(i)}(t) = \qty((\vb{\phi}_l^{(i)} e^{-i 2\pi f_l t}) \circledast \textbf{q}) (t) =\int_{\Delta T} \int_{\Omega} \qty( \vb{\phi}_l^{(i)}(x))^{*}\textbf{W}(x)\textbf{q}(x,t+\tau)w(\tau) e^{-i 2\pi f_l\tau}\dd x \dd \tau,
     \label{eq exp_conv}
    \end{equation}
where $\circledast$ indicates the convolution between the time evolving SPOD mode and the data, that takes into the account the windowing function, $w(\tau)$, and the weight matrix $\vb{W}$. In practise this convolution is computed by expanding the SPOD mode in time as  $\vb{\phi}_l^{(i)} e^{-i 2\pi f_l t}$ and convolving it over the data one snapshot at a time.   In this step, we leverage the orthogonality property of the SPOD mode in the space-time inner product, which allows us to compute the expansion coefficient one at a time.   If the SPOD was computed using an overlap of $n_{\rm fft}$-1, then the expansion coefficients, $\vb{a}_l^{(i)}$, obtained from equation (\ref{eq exp_conv}) and equation (\ref{eq 15}) are mathematically identical.  Here, the underlying idea is to apply the continuously-discrete convolution integral to the SPOD mode computed with a significantly lower overlap to make it computationally feasible. We confirmed that the frequency-time diagrams of the expansion coefficients obtained using equation (\ref{eq exp_conv}) for an overlap of $50\%$ are virtually indistinguishable to those obtained from equation (\ref{eq 15}) for an $n_{\rm{ovlp}} = n_\mathrm{fft}-1$ (shown in the appendix \ref{appendix:ft-analysis  frequncy and convolution comparison }, figure \ref{fig con_freq}).  This is to be expected since the convergence of the SPOD modes does not improve significantly for overlap over $50\%$. In practice, the convolution integral in equation (\ref{eq exp_conv}) is most efficiently computed using the FFT.

Frequency-time diagrams of the expansion coefficients for the convolution approach are shown in figure \ref{fig 13}. Figure \ref{fig 13}($a$), and ($b$) show the contribution of the leading 10 modes, and the leading mode, respectively. These frequency-time diagrams appear less detailed compared to the frequency-time diagrams of the time-domain approach. In the context of figure \ref{fig 14}, however, we will show that figures \ref{fig 12} and \ref{fig 13} basically contain the same information. For now, it is sufficient to note that the convolution and time-domain approaches detect the same trends. Take as an example, the high energy events occurring at low frequency in the time ranges, $550 \lesssim t \lesssim 590$ and $1370 \lesssim t \lesssim 1400$, in the spectrograms of figure \ref{fig 12}($b$) and \ref{fig 13}($b$).

\begin{figure}
\centering
{\includegraphics[trim={0cm 2.25cm 0cm 1.9cm},clip, width=1.0\textwidth]{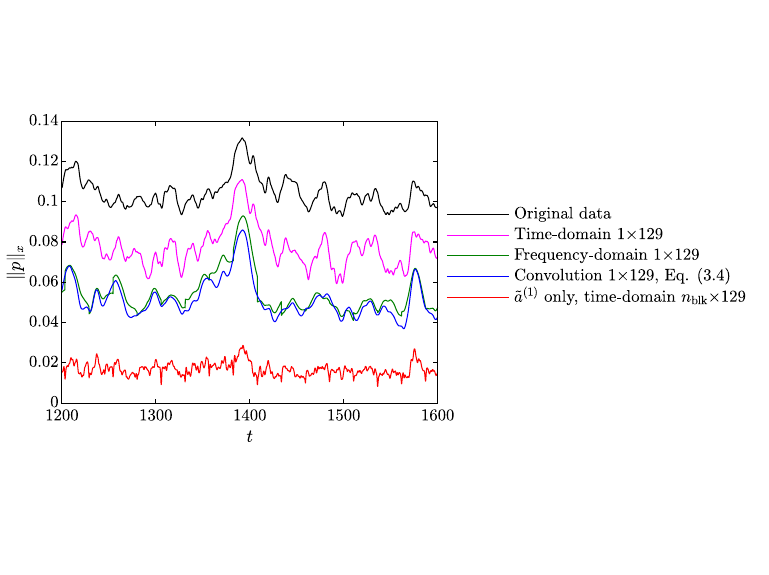}}
\caption{Temporal evolution of the pressure 2-norm in the vicinity of its global maximum at $t=1380$. The convolution approach, based on equation (\ref{eq exp_conv}), is compared to the time- and frequency-domain approaches, previously shown in figure \ref{fig 3}($b$,$d$). The 1$\times$129-mode reconstructions are shown. 
}
\label{fig 12a}
\end{figure}

\subsubsection{Comparison of methods and interpretation of results}
Next, we investigate if the high similarity between the instantaneous flow field and the modes indicates high energy. Figure \ref{fig 12a} shows the temporal evolution of the pressure 2-norm for the original data, low-dimensional 1$\times$129-mode reconstructions using the time-domain, frequency-domain and the convolution approaches. The pressure 2-norm of the data is highly underpredicted by the time-domain approach that uses a full basis ($n_{\rm blk}\times$129), but is able to capture the major trends of the original data. The time-domain reconstruction performed using a modal basis of 1$\times$129 is also shown for comparison. This  accurately follows the spatial norm of the original data, except for an offset, similar to figure \ref{fig 3}($d$). The frequency-domain curve also shows a similar trend. In addition, the spatial norm of the 1$\times$129-mode reconstruction using the convolution approach is shown. It follows the trend of the original data and attains its global maximum at the same time instant. All curves in figure \ref{fig 12a} peak at the time of maximum instantaneous energy, previously indicated in figure \ref{fig 13}($b$). This indicates that a high similarity between the instantaneous flow field and the leading mode implies high overall energy. We highlight that this finding is not self-evident as the leading SPOD mode represents the most energetic flow structure in a purely \emph{statistical} sense. The important physical insight hence is that the intermittent occurrence of large-scale coherent structures is directly associated with high-energy events.

\begin{figure}
\centering
{\includegraphics[trim={0cm 0.1cm 0cm 0.3cm},clip, width=1.0\textwidth]{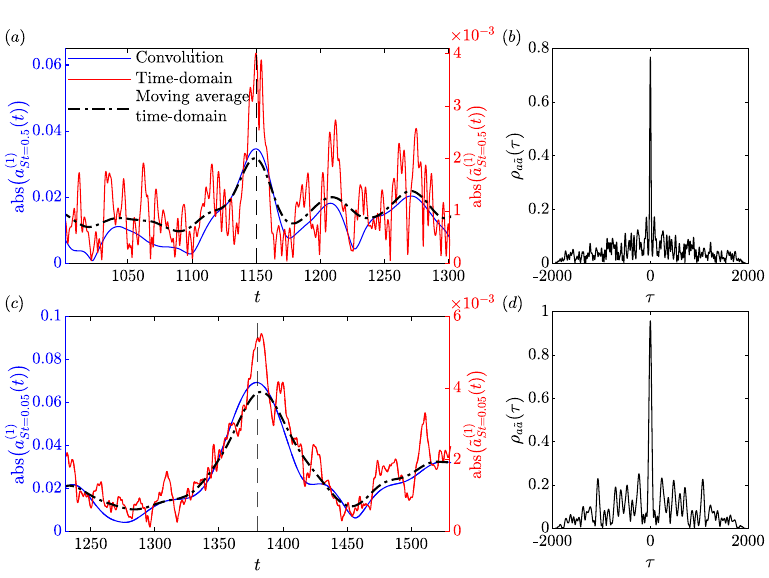}}
\caption{Oblique projection and convolution-based expansion coefficients of the leading mode at $\St=0.5$ (top) and $\St=0.05$ (bottom): ($a$,$c$) time traces in the vicinity of their global maximum (black dashed line); ($b$,$d$) cross-correlation of the expansion coefficients obtained using the two different approaches. The black dash-dotted line denotes the weighted moving mean of the time-domain expansion coefficient. The moving mean uses as weights the same Hamming windowing function as the SPOD.}
\label{fig 14}
\end{figure}

To understand the qualitative differences of the frequency-time diagrams in figure \ref{fig 12} and \ref{fig 13}, we now look at the expansion coefficients of the two approaches. Figure \ref{fig 14} compares the expansion coefficient of the leading mode, obtained by the time-domain and convolution approaches. As an example, the expansion coefficient at $St \approx 0.50$ is shown in figure \ref{fig 14}($a$). It is centered around its global maximum, in the time interval $1000 \le t \le 1300$. We observe that the time traces of the expansion coefficients obtained from the two approaches show similar trends. In particular, the local peaks occur at similar locations, with both curves exhibiting the global maximum at $t=1150$ (black dashed line). Compared to the time-domain approach, the convolution curve is smoother and resembles a moving average of the time-domain curve (black dash-doted line). For optimal comparison with the convolution approach, the moving average is computed by weighting the time-domain curve by the Hamming window in equation (\ref{eq 6}) and averaging it 256 points. To further quantify the relation between the expansion coefficients of the two approaches, we show the cross-correlation coefficient in figure \ref{fig 14}($b$).  The cross-correlation coefficient confirms the observation that the expansion coefficients obtained from the two approaches are similar, by demonstrating a cross-correlation coefficient of 0.77 at $0$-time lag ($\tau = 0$). We have confirmed that this correspondence holds in general. In figure \ref{fig 14}($c$) and \ref{fig 14}($d$), for example, the same trends are observed for the expansion coefficients at the lower frequency of $\St=0.05$ over the time interval previously shown in figure \ref{fig 12a} above. From figure \ref{fig 14}, we infer that the intermittency of the coherent structures can be captured using both approaches. 
    
\begin{figure}
\centering
{\includegraphics[width=1.0\textwidth]{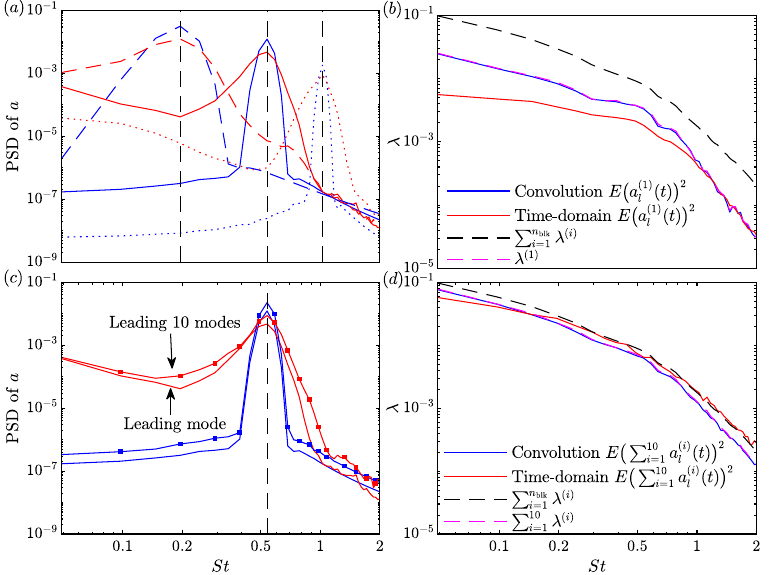}}
\caption{ Spectral analysis of the expansion coefficients for the time-domain and convolution approaches: ($a$) PSD of the individual expansion coefficients of the leading SPOD mode at $\St = 0.2$, $0.5$ and $1.0$; ($b$) expected values of the expansion coefficients of the leading mode; ($c$) PSD of the expansion coefficients of the leading mode and sum of the first 10 leading modes (line with squares) at $\St = 0.5$; ($d$) expected value of the expansion coefficients for the sum of the first 10 modes. In ($b$) and ($d$), the leading, sum of leading 10, and sum of all eigenvalues are shown for comparison.}
\label{fig 15}
\end{figure}

After establishing the correspondence between the time-domain and convolution approaches, we now compare the  spectral characteristics of the two approaches. The PSDs of the expansion coefficients associated with the leading mode at $\St=0.2$, 0.5, and 1.0, for the time and convolution-domain approaches are shown in figure \ref{fig 15} ($a$). As expected the PSDs peak at the frequency of the corresponding mode for both approaches. The expansion coefficients computed in the convolution approach exhibit a much narrower peak than those in the time-domain approach. Note that, we cannot expect a sharp spectral peak even for the convolution approach due to spectral leakage and the modulation of the wave amplitude as seen in figure \ref{fig 14}. Figure \ref{fig 15}($c$) compares the PSDs of the leading mode and the 10 leading modes at $\St = 0.5$, which underlines the dominance of the leading mode at this particular frequency.
    The expansion coefficients are presented in terms of the spectral energy content in figure \ref{fig 15}($b$) and ($d$). As the expansion coefficients are uncorrelated, and their expected value is the equal to SPOD modal energy, 
    \begin{equation}
    E\Bqty{\vb{a}_f^{(i)}\vb{a}_f^{(j)}}=\lambda_f^{(i)} \delta_{ij},
    \label{eq 3.10}
    \end{equation}
    it is expected that the convolution approach accurately approximates the eigenvalue spectrum. The blue and magenta lines are almost coincident in figure \ref{fig 15}($b$,$d$), thus confirming this conjecture.  For the time-domain approach, the expected value for the sum of first 10 modes approximates the total integral PSD for all but the low frequencies $\St \ge 0.2$ in figure \ref{fig 15}($d$). On considering only the first eigenvalue, the time-domain approach under predicts the eigenvalue spectrum for $\St \le 1.0$, (figure \ref{fig 15}($b$)). Note that, this observation does not contradict the observations made in context of  figure \ref{fig 3}, as the expansion coefficients are obtained from a full basis here, but from a smaller basis in figure \ref{fig 3}($d$).

\begin{figure}
\centering
{\includegraphics[trim={0cm 1.05cm 0cm 2.05cm},clip, width=1.0\textwidth]{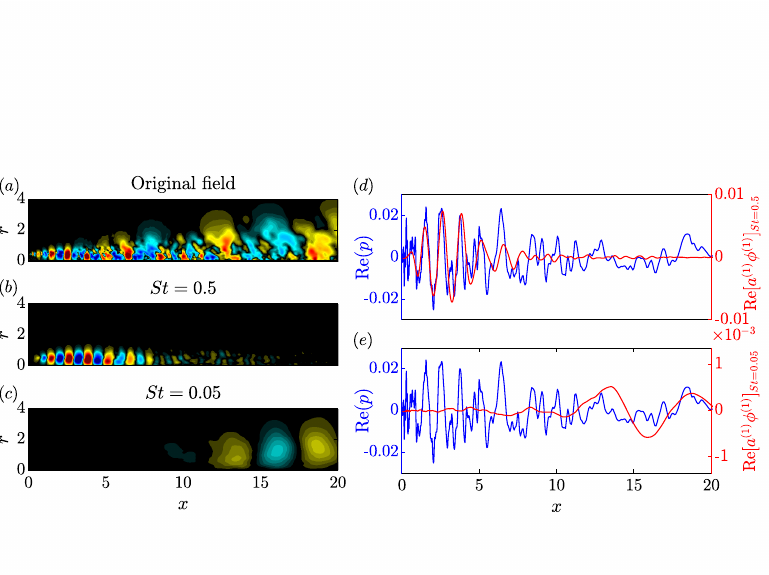}}
\caption{Contributions of individual modes to the reconstructed flow field: ($a$) instantaneous pressure field; ($b$) contribution of ${\phi}^{(1)}$ for $St$=0.5, and ($c$) for $St$=0.05; ($d$,$e$) comparison of the real part of the pressure field along the lipline ($r=0.5$) for $St$=0.5 and $St$=0.05, respectively. The instantaneous flow field at $t=$1150 (see figure \ref{fig 14}) is shown. Contours in ($a,b,c$) are reported on the same color axis.}
\label{fig 16}
\end{figure}

After examining the properties of the expansion coefficients, we focus on the spatial composition of the reconstructed flow field in terms of contributions from individual modes in figure \ref{fig 16}. The time instant of high energy, previously marked in figure \ref{fig 14}, is chosen as an example. The contribution of the leading SPOD modes at two representative frequencies, $\St=0.5$, and 0.05, to the original flow field is shown in figure \ref{fig 16}. Figure \ref{fig 16}($b$) and ($c$) show the modes weighted by their expansion coefficient, $a^{(1)}\phi^{(1)}$ at the corresponding time instant and these two frequencies.  Close to the nozzle exit the LES flow field clearly exhibits a KH-type instability wave. The leading mode at $\St=0.5$, closely resembles this structure. Similarly, the dominant wave pattern with a large wavelength ($\approx 5$) observed in the flow field, is represented by the mode at $\St=0.05$ in a location downstream of the potential core ($12\lesssim x\lesssim 20$). The real part of the pressure field along the lipline ($r=0.5$) is compared with the contributions of the leading modes at $\St=0.5$, and $\St=0.05$ in figure \ref{fig 16}($d$), and ($e$), respectively.  It can be seen that the phases of the pressure field and the mode at $\St=0.5$ are aligned in the region where the mode attains its maximum.   Weaker but similar kind of phase alignment is also observed for low frequency, $\St=0.05$, despite the disturbed nature of the wavepacket in the region, $12\lesssim x \lesssim 20$. For the flow field at the current time instant, this also explains that the contribution of the leading mode at $\St=0.05$ is lower than $\St=0.5$, where the KH wavepacket dominates the flow field. These observations confirm  that the maximum in the frequency-time diagrams indicate close resemblance of the instantaneous flow field with the corresponding SPOD modes. Since the SPOD modes are the most energetic structures it is not surprising that the maxima in the frequency-time diagrams indicate the intervals of high energy. Furthermore, as the leading SPOD modes are spatially coherent and contain the most energy at each frequency, we infer that SPOD-based frequency-time analysis can be used to gauge the intermittency of large-scale coherent structures. Here, for brevity, only the convolution approach is shown, but we note that these conclusions also hold for the time-domain approach. 

\subsection{Prewhitening} \label{appendix prewhitening}

\begin{figure}
\centering
{\includegraphics[trim={0cm 2.8cm 0cm 0},clip, width=1.00\textwidth]{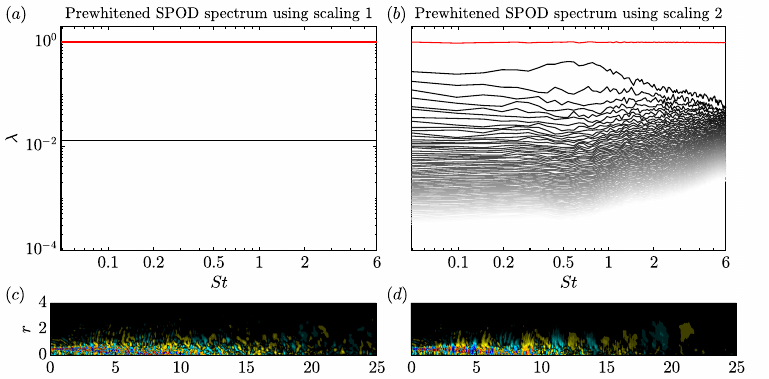}}
\caption{SPOD-based prewhitening: ($a,b$) rescaled SPOD spectra; ($c,d$), instantaneous pressure fields using definitions (\ref{eq 3.11a}) and (\ref{eq 3.11b}), respectively. Red lines indicate the sum of all eigenvalues, that is, the integrated PSD. All lines in ($a$) collapse due to the scaling. The original SPOD eigenvalue spectrum can be seen in figure \ref{fig 1}($a$).}
\label{fig prewhitening}
\end{figure}

As mentioned in \S\ref{introduction}, prewhitening is a filtering operation that results in a flat power spectrum, and is commonly used for trend detection in  atmospheric and geophysical applications. The goal hence is to use SPOD to scale the data to have the same energy at all frequencies. We propose to achieve this goal by re-scaling the expansion coefficients of the reconstruction in the frequency-domain. The frequency-domain is chosen for the same reasons as for denoising in \S\ref{denoising}. We leverage the fact that expansion coefficients are uncorrelated, and that their expected value is the equal to SPOD modal energy, 
    \begin{equation}
    E\Bqty{\vb{a}_f^{(i)}\vb{a}_f^{(j)}}=\lambda_f^{(i)} \delta_{ij}.
    \label{eq 3.10}
    \end{equation}
 We propose two different scalings,   
    \begin{eqnarray}
         \frac{\sqrt{\sum\limits_f \sum\limits_i \lambda_f^{(i)}}}{\sqrt{\lambda_f^{(i)}}}&\vb{a}_f^{(i)}&\quad\text{(scaling 1), and} \label{eq 3.11a} \\
          \frac{1}{\sqrt{\sum\limits_i\lambda_f^{(i)}}}&\vb{a}_f^{(i)}&\quad\text{(scaling 2)} \label{eq 3.11b}
    \end{eqnarray}
to scale the integral mode energy, that is, the sum of all eigenvalues, to one at each frequency. Both methods achieve this goal, but result in different relative scalings of individual SPOD modes. Figure \ref{fig prewhitening}($a$) and \ref{fig prewhitening}($b$) show the effect of the two scalings on the SPOD eigenvalue spectrum. The reconstructed flow fields obtained from the expansion coefficients scaled using equations (\ref{eq 3.11a}) and (\ref{eq 3.11b}) are shown in \ref{fig prewhitening}($c$) and \ref{fig prewhitening}($d$), respectively. Note that equation (\ref{eq 3.11a}) collapses all eigenvalues in figure \ref{fig prewhitening}($a$) to the same value. Scaling 2, on the contrary, preserves both the mode hierarchy and the relative energy content. In comparison to the original flow field shown in figure \ref{fig 5}($a$), prewhitening emphasizes high-frequency structures in the shear-layer, whereas it de-emphasizes the highly energetic large-scale structures associated with low frequencies downstream of the potential core. This portrayal of the flow field might appear unfamiliar; we emphasize that the objective of prewhitening is not physical interpretation, but pattern identification. Here, for example, the prewhitened pressure fields bring to light the trapped acoustic modes in the potential core. These modes have only recently been described in detail \citep{towne2017acoustic, schmidt2017wavepackets}. Previously, they remained largely unnoticed in the analysis of jet data because of their low energy content. An important difference between SPOD-based prewhitening and classical local, point-wise prewhitening techniques is that the SPOD-based approach preserves the spatial coherence of the flow structures identified by the SPOD modes.

\section{Summary and Conclusions} \label{sec:conclusion}

Different applications of SPOD including low-rank reconstruction, denoising, prewhitening, and frequency-time analysis are demonstrated on the example of LES data of a turbulent jet. A fundamental building block for these applications is the capability to reconstruct the original data from the SPOD. In the frequency domain, this can be accomplished by inverting the SPOD problem \citep[see, e.g.,][]{citriniti2000reconstruction}. We demonstrate that this inversion can be computed either directly in the frequency-domain, or using a convolution-based strategy. The latter approach becomes a necessity in the context of frequency-time analysis, where the corresponding SPOD problem becomes intractable. As an alternative to frequency-domain reconstruction, we introduce a time-domain approach that is based on the oblique projection of the data onto the SPOD modes.

First, we show the complete recovery of the data using all modes and compare to low-dimensional reconstructions.  The low-dimensional reconstructions from both approaches accurately capture the integral energy of the segmented data (in the space-time norm). However, the time-varying dynamics (in the purely spatial norm) are only captured by the low-dimensional reconstructions in the time-domain. For a fixed number of modes, the time-domain approach captures more of the energy (in both norms). On the downside, the association of the SPOD modes with a single frequency is lost. Instantaneous pressure fields reconstructed in the frequency-domain, on the contrary, preserve this monochromatic property of the SPOD modes, but may lack the finer details of the flow field reconstructions in the time-domain.  The main advantage of the frequency-domain approach is that it conserves the orthogonality property and the frequency-mode correspondence of the SPOD. The main advantage of the time-domain approach is its optimality in reconstructing the instantaneous flow field with the least possible number of modes.

After establishing the advantages and disadvantages of both approaches, we demonstrate SPOD-based denoising as an application of the frequency-domain approach. As expected, noise is mainly captured by higher SPOD modes at low frequencies and all modes at high frequencies. As a best practice, we propose a hard threshold above the noise floor that is identified from the SPOD eigenvalue spectrum. Significant noise reduction is achieved. At the same time, a substantial amount of energy of the original flow field is retained. Compared to a standard low-pass filter, SPOD-based denoising has the additional advantage of drastic space savings.

Finally, we demonstrate how SPOD-based frequency-time analysis can be used to analyse the intermittency of turbulent flows.  Established means of frequency-time analysis such as wavelets transforms are signal-processing techniques that are applied to 1-D time signals. The alternative, SPOD-based approach demonstrated here, provides a global perspective in which spectrograms characterize the temporal evolution of the spatially coherent flow structures represented by the SPOD modes. The SPOD-based frequency-time analysis requires the computation of time-varying expansion coefficients at each time instant, and is computationally intractable in the frequency-domain. This problem is mitigated by the convolution-based strategy, which is mathematically equivalent in the limit of the intractable continuously-discrete (in time) SPOD problem.  This convolution-based approach is compared to the projection-based approach in the time domain. The expansion coefficients calculated from both methods show similar trends. We further demonstrate that a moving average of the spectrogram obtained via oblique projection resembles the spectrogram obtained from the convolution approach. For consistency, the moving time average is directly based on the SPOD windowing. The main advantage of the frequency-domain, and therefore the convolution approach, is that it retains the  orthogonality property and mode-frequency correspondence of the SPOD. The frequency-time analysis of the jet data confirms the highly intermittent nature of this turbulent flow. In accordance with the SPOD eigenvalue spectrum, it is found that most of the energy is concentrated at low frequencies, $St \lesssim 0.2$. A comparison of the total flow energy as a function of time with the spectrograms shows that high energy events are directly linked to the presence of flow structures resembling the leading SPOD modes. We highlight that this behaviour is not necessarily expected as the SPOD modes represent the most energetic structures only in a \emph{statistical} sense. From previous work \citep{schmidt2018spectral}, it is well-known that SPOD modes often isolate certain, prevailing physical phenomena. SPOD-based frequency-time analysis hence provides additional physical insight by indicating time intervals during which a particular mechanism is active.

Based on the results, we recommend the use of the time-domain approach for low-rank reconstruction of individual snapshots, and the frequency-domain approach for denoising and frequency-time analysis. For the latter application the proposed convolution strategy facilitates efficient computation of the time-continuous expansion coefficients.\newline

\noindent \textbf{Acknowledgements}
\newline We gratefully acknowledge support from Office of Naval Research grant N00014-20-1-2311. The authors would like to thank the anonymous reviewers for their insightful comments. In particular, we thank the first reviewer for the suggestion regarding time-continuity. 

\noindent \textbf{Declaration of interests.}  The authors report no conflict of interest.

\appendix

\section{Effect of different parameters on the flow field reconstruction} \label{appendix 1}

The segmentation of the data is a crucial step in spectral estimation. Following the original work by \citep{welch1967use}, we use an overlap of $50\%$ between blocks to minimize the variance of the spectral estimate, see \S \ref{sec:results} above. The use of overlapping segments, however, results in an ambiguity for the reconstruction, which we may compute as/from:
\begin{enumerate}
    \item the left (previous) block,
    \item the right (following) block,
    \item the average of the left and right reconstructions,
    \item either the left or right reconstruction based on the higher windowing weight,
    \item the average of the left and right reconstructions weighted by the relative value of hamming window, 
    \item the average of the left and right reconstructions weighted by the relative distance to the centers of the overlapping blocks.
\end{enumerate}

\begin{figure}
\centering
{\includegraphics[trim={0.0cm 3.6cm 0.0cm 1.3cm },clip,width=1.0\textwidth]{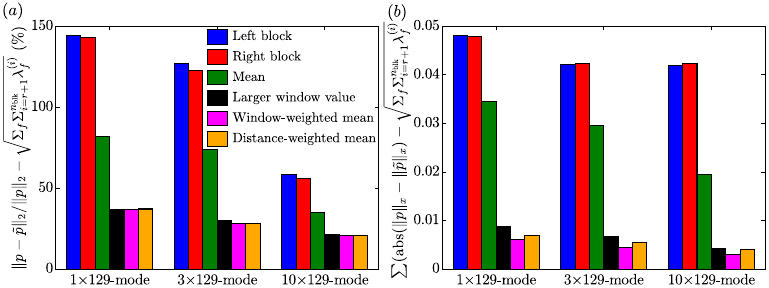}}
\caption{ Errors of different ways to compute the  frequency-domain reconstruction for 1$\times$129, 3$\times$129 and 10$\times$129 modes: ($a$) 2-norm; ($b$) spatial norm (equation (\ref{eq 2})). Analogous to figure \ref{fig 4}, the residual energy of the truncated modes is subtracted for comparison.}
\label{fig A1}
\end{figure}

Figure \ref{fig A1} compares the errors of low-dimensional reconstructions using 1$\times$129,  3$\times$129, and 10$\times$129 modes for all the six possibilities.  It is found that the reconstruction based on the window-weighted average of the left and right reconstructions produces the smallest error. Based on this finding, this option is used for the frequency-domain reconstruction throughout the paper. Using the window-weighted average approach, the $i$-th snapshot is reconstructed as,  
\begin{equation}
    \vb{q}_{i}  \approx \frac{\vb{q}_{j}^{(k)}w(j)+\vb{q}_{j-n_{\rm{ovlp}}}^{(k+1)}w\pqty{\lvert j-n_{\rm{ovlp}}\rvert}}{w(j)+w\pqty{\lvert j-n_{\rm{ovlp}}\rvert}},  
    \end{equation} where $j$= $i - (k-1)(n_{\rm{fft}}-n_{\rm{ovlp}})$, $i\in [1,n_t]$, $j\in[1,n_{\rm{fft}}]$ and $k\in[1,n_{\rm{blk}}]$. From figure \ref{fig 3} (in particular \ref{fig 3}($b$)), it becomes apparent that these distinctions only matter for truncated series reconstructions; full-dimensional reconstructions are generally accurate. 


\begin{figure}
\centering
{\includegraphics[width=1.0\textwidth]{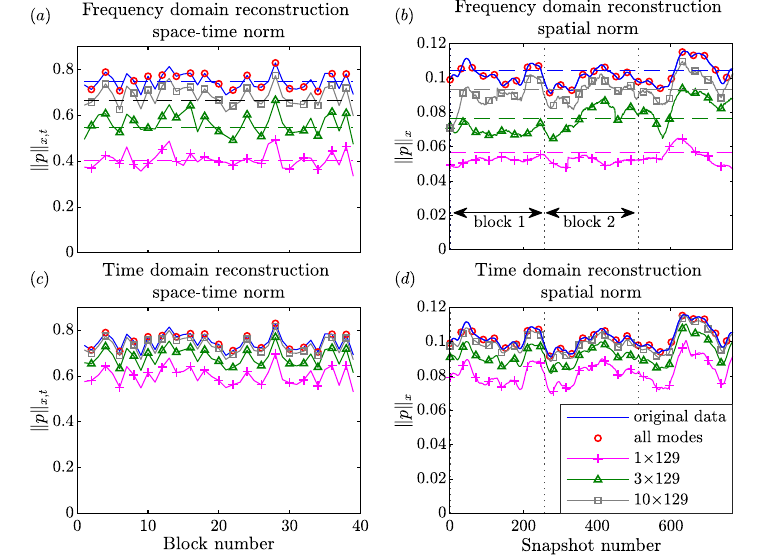}}
\caption{Low-dimensional reconstruction using a rectangular window: frequency-domain reconstruction ($a,b$), and time-domain reconstruction on the ($c,d$) in terms of the space-time norm ($a,c$) and the spatial norm ($b,d$). The original data (blue lines) is compared to the full reconstructions using all modes, and reconstructions using 10$\times$129, 3$\times$129, and 1$\times$129 modes. Summed SPOD mode energies are shown as dashed lines. Vertical dotted black lines in ($b$,$d$) indicate the non-overlapping blocks. This figure is identical to figure \ref{fig 3},  but for a rectangular window and no overlap.}
\label{fig A2}
\end{figure}

The sudden jumps in the local energy of the reconstruction  observed in figure \ref{fig 3} are a windowing effect. We demonstrate this by comparison with reconstruction using rectangular windows and no overlap in figure \ref{fig A2}. The low-dimensional reconstructions using 1$\times$129, 3$\times$129, and 10$\times$129 modes, and all modes are shown. Many observations made in the context of figure \ref{fig 3}, also hold here: the dimension of the modal bases is directly proportional to its ability to capture the pressure norm of the data, and for a fixed number of modes, the time-domain results in a better approximation of the data than the frequency-domain approach. The most notable difference can be seen between figure \ref{fig A2}($b$) and \ref{fig 3}($b$). The windowing effect in the frequency-domain reconstructions is absent if the rectangular window is used. Note in particular the difference during the first few snapshots and near the locations of switching from one block to another (vertical dotted black lines). Despite this advantage in the context of frequency-domain reconstructions for small $n_{\rm{modes}}$, rectangular windowing is generally not recommended because of spectral leakage \citep{schmidt2020guide}.

\begin{figure}
\centering
{\includegraphics[trim={0cm 2.75cm 0cm 1.4cm},clip,width=1.0\textwidth]{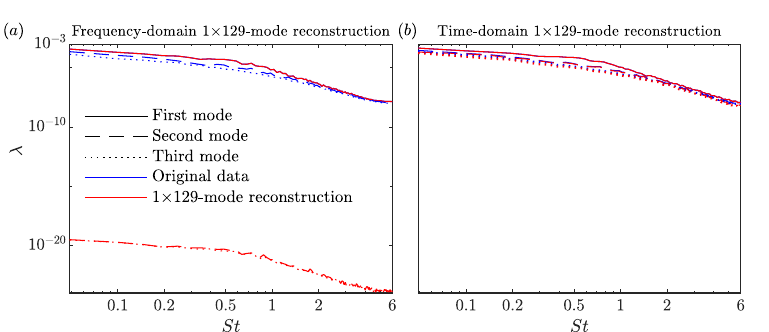}}
\caption{ Comparison between SPOD eigenvalue spectra of the original data (blue lines) and 1$\times$129-mode reconstructions (red lines): ($a$) frequency-domain; ($b$) time-domain. Solid, dashed, and dotted lines denote the first, second, and third modes, respectively. This figure is identical to figure \ref{fig 6},  but for a rectangular window and no overlap.}
\label{fig A3}
\end{figure}

Analogous to figure \ref{fig 6}, we report in figure \ref{fig A3} the SPOD eigenspectra of the 1$\times$129-mode frequency and time-domain reconstructions for a rectangular window and $n_{\rm{ovlp}}=0$. The SPOD eigenvalue spectra of the full data is also shown for comparison. Only the leading three eigenvalues are shown for clarity. The leading eigenvalue of the frequency-domain and time-domain reconstructions are indistinguishable from the leading eigenvalues of the full data. For the frequency-domain reconstruction,  the higher eigenvalue spectra are zero to machine precision, as expected. This indicates that the windowing-effect causes the elevation of the higher eigenvalue spectra in figure \ref{fig 6}($a$). The time-domain reconstruction, on the other hand, is able predict the higher eigenvalues as explained in the context of \ref{fig 6}($b$). This implies that the time-domain reconstruction is much less sensitive to the choice of windowing function.

  
\section{Projection-based frequency-time analysis: effect of choice of basis and correspondence to convolution-based approach}  \label{appendix:frequnecy-time analysis }

  \begin{figure}
\centering
{\includegraphics[trim={0cm 1.4cm 0cm 1.9cm},clip,width=1.0\textwidth]{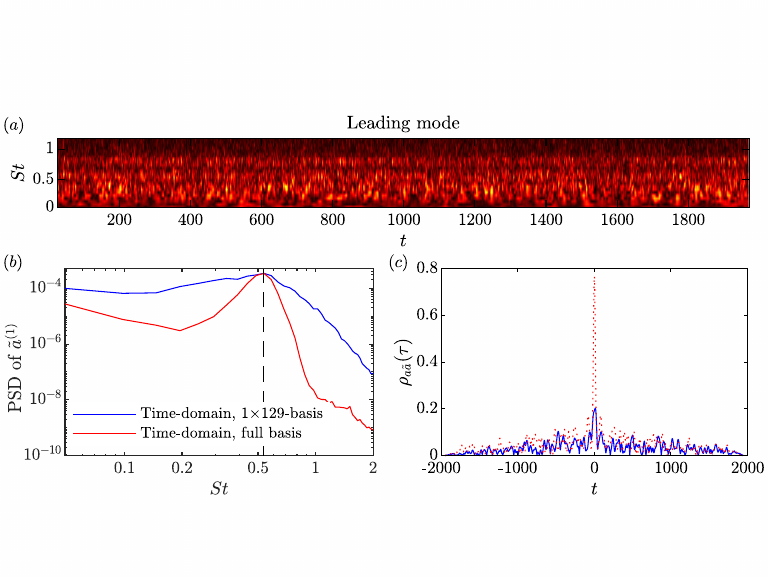}}
\caption{SPOD-based frequency-time analysis obtained using the time-domain approach with 1$\times$129-modal basis: ($a$) frequency-time diagram of the leading mode at each frequency; ($b$) PSD of the individual expansion coefficients of the leading SPOD mode at $\St=0.5$ for the time-domain approach with the 1$\times$129 basis (blue line) and $n_{\rm{blk}}\times$129 basis (red line); ($c$) cross-correlation of the expansion coefficient in the time-domain approach using the 1$\times$129 basis (blue line) and the $n_{\rm{blk}}\times$129 basis (red dotted line)  with the convolution approach.}
\label{fig A5}
\end{figure}

Oblique projection-based frequency time analyses is dependent on the choice of modal basis. The two obvious choices of bases for the projection are
\begin{enumerate}
    \item $n_{\rm{modes}}\times$129 modes, i.e., only those SPOD modes used in the analysis, 
    \begin{equation*}
\tilde{\vb*{\Phi}}= \bqty{\vb{\boldsymbol{\phi}}_{1}^{(1)}, \vb{\boldsymbol{\phi}}_{2}^{(1)}, \cdots, \vb{\boldsymbol{\phi}}_{n_{\rm fft}}^{(1)}}
\label{eq C1}, \quad \text{or}
\end{equation*}
    \item $n_{\rm{blk}}\times$129 modes, i.e., all available SPOD modes,
  \begin{equation*}
\tilde{\vb*{\Phi}}= \bqty{\vb{\boldsymbol{\phi}}_{1}^{(1)}, \vb{\boldsymbol{\phi}}_{1}^{(2)}, \cdots , \vb{\boldsymbol{\phi}}_{1}^{(n_{\rm blk})}, \vb{\boldsymbol{\phi}}_{2}^{(1)}, \vb{\boldsymbol{\phi}}_{2}^{(2)}, \cdots , \vb{\boldsymbol{\phi}}_{2}^{(n_{\rm blk})}, \cdots, \vb{\boldsymbol{\phi}}_{n_{\rm fft}}^{(1)}, \vb{\boldsymbol{\phi}}_{n_{\rm fft}}^{(2)}, \cdots , \vb{\boldsymbol{\phi}}_{n_{\rm fft}}^{(n_{\rm blk})} }.
\label{eq C2}
\end{equation*}
\end{enumerate}
Due to the non-orthogonality of these modes in the spatial norm, these two choices will result in different outcomes.  Shown in figure \ref{fig A5}($a$) is the frequency-time diagram for $n_{\rm{modes}}$=1, that is a $1\times$129-mode basis containing only the leading mode at each frequency. A fundamentally different behaviour from that in figure \ref{fig 12} is observed. The diagram exhibits a banded structure, and, in contrast to the reference diagram based on all SPOD modes, the majority of maxima is not found in the low-frequency regime, $St \lesssim 0.2$. To understand this difference, the PSD of the expansion coefficient associated with the leading mode at $\St=0.5$ is shown in figure \ref{fig A5}($b$). The expansion coefficient computed with the $1\times$129-modal basis exhibits a much broader peak than the one computed with full basis. This behaviour indicates a loss of the mode-frequency correspondence for the heavily truncated basis.  Next, both approaches are compared by taking the convolution-based expansion coefficient as the reference. The cross-correlation of the expansion coefficients from both approaches with the reference signal from the convolution approach are shown in figure \ref{fig A5}($c$). The expansion coefficient computed using the $1\times$129-modal basis exhibits a much lower correlation with the reference. We conclude from this analysis that the time-domain approach should be conducted using all SPOD modes, as it yields a more accurate description of the intermittency of the coherent structure represented by the SPOD modes.   

\begin{figure}
\centering
{\includegraphics[trim={0cm 3.45cm 0cm 1.75cm},clip, width=1.0\textwidth]{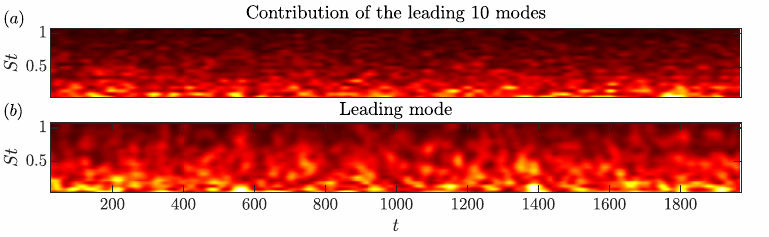}}
\caption{SPOD-based frequency-time diagrams obtained using the moving mean of the time-domain oblique projection based approach: ($a$) first 10 modes at each frequency; ($b$) leading mode at each frequency. The moving mean uses as weights the same Hamming windowing function as the SPOD.}
\label{fig A6}
\end{figure}

In figure \ref{fig 14}, we demonstrated that the expansion coefficients computed from a moving average of the time-domain approach resemble those from the convolution approach. For further evidence, we show in figure \ref{fig A6} the frequency-time diagrams obtained by taking the moving mean, at each frequency, of the time-domain diagram previously shown in figure \ref{fig 12}. The outcome should be compared to the frequency-time diagrams obtained using the convolution approach, i.e., figure \ref{fig 13}. It is observed that the frequency-time diagrams are qualitatively very similar.  The effect of taking the moving average is mainly visible at higher frequencies, where it leads to minor loss of detail. For qualitative flow analysis, we hence conclude that the moving average of the time-domain approach can well be used to approximate the computationally much more involved convolution approach.

\section{Frequency-domain approach based frequency-time analysis: effect of overlap and correspondence to convolution-based approach}  \label{appendix:ft-analysis  frequncy and convolution comparison }

\begin{figure}
\centering
{\includegraphics[trim={0cm 0 0cm 0cm },clip,width=1.0\textwidth]{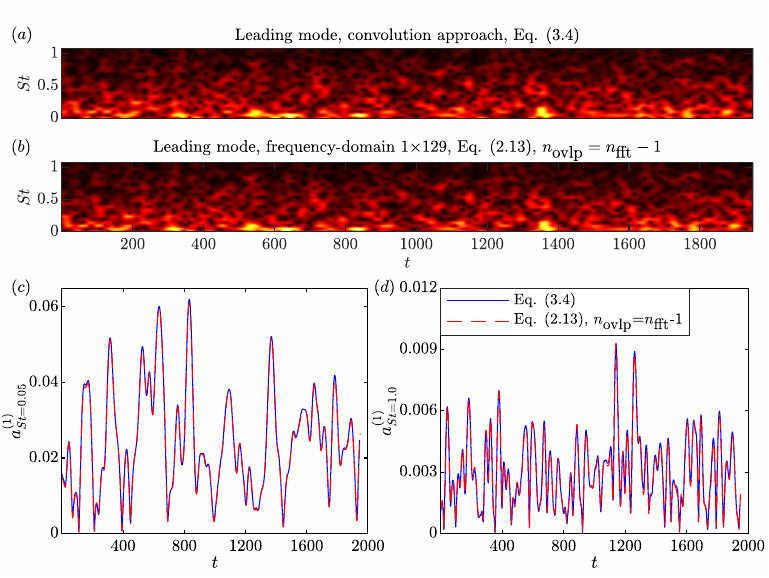}}
\caption{Comparison between two strategies for SPOD-based frequency-time analysis: frequency-time diagrams for ($a$)  the convolution approach (SPOD modes precomputed with 50\%, overlap), and ($b$) the frequency-domain approach using $n_{\rm{ovlp}}=n_\textrm{fft}-1$; expansion coefficients for ($c$)  $\St=0.05$, and ($d$) $\St=1.0$.}
\label{fig con_freq}
\end{figure}

Figure \ref{fig con_freq} demonstrates the similarity between the frequency-time diagrams obtained using the convolution method and the frequency-domain approach. For the convolution method, equation (\ref{eq exp_conv}), a basis of precomputed SPOD modes with 50\% overlap was used. To obtain time-continuous expansion coefficients using the frequency-domain approach, we require $n_{\rm{ovlp}}=n_\textrm{fft}-1$. As this is computationally intractable, the data (only here) was reduced to every third grid point in the streamwise and radial directions. The time traces of the individual expansion coefficients for $\St=0.05$ and 1.0 are reported in \ref{fig con_freq}($c$,$d$).

\bibliographystyle{jfm}
\bibliography{jfm}

\begin{thebibliography}{87}
\expandafter\ifx\csname natexlab\endcsname\relax\def\natexlab#1{#1}\fi
\def\au#1{#1} \def\ed#1{#1} \def\yr#1{#1}\def\at#1{#1}\def\jt#1{\textit{#1}}
  \def\bt#1{#1}\def\bvol#1{\textbf{#1}} \def\vol#1{#1} \def\pg#1{#1}
  \def\publ#1{#1}\def\arxiv#1{#1}\def\org#1{#1}\def\st#1{\textit{#1}}

\bibitem[Arndt {\em et~al.\/}(1997)Arndt, Long \& Glauser]{arndt1997proper}
{\sc \au{Arndt, R. E.~A.}, \au{Long, D.~F.} \& \au{Glauser, M.~N.}} \yr{1997}
  \at{The proper orthogonal decomposition of pressure fluctuations surrounding
  a turbulent jet}.  \jt{Journal of Fluid Mechanics}  \bvol{340},  \pg{1--33}.

\bibitem[Aubry(1991)]{aubry1991hidden}
{\sc \au{Aubry, N.}} \yr{1991}  \at{On the hidden beauty of the proper
  orthogonal decomposition}.  \jt{Theoretical and Computational Fluid Dynamics}
   \bvol{2}~(5-6),  \pg{339--352}.

\bibitem[Bale {\em et~al.\/}(2005)Bale, Kellogg, Mozer, Horbury \&
  Reme]{bale2005measurement}
{\sc \au{Bale, S.~D.}, \au{Kellogg, P.~J.}, \au{Mozer, F.~S.}, \au{Horbury,
  T.~S.} \& \au{Reme, H.}} \yr{2005}  \at{Measurement of the electric
  fluctuation spectrum of magnetohydrodynamic turbulence}.  \jt{Physical Review
  Letters}  \bvol{94}~(21),  \pg{215002}.

\bibitem[Boashash(1988)]{boashash1988note}
{\sc \au{Boashash, B.}} \yr{1988}  \at{Note on the use of the wigner
  distribution for time-frequency signal analysis}.  \jt{IEEE Transactions on
  Acoustics, Speech, and Signal Processing}  \bvol{36}~(9),  \pg{1518--1521}.

\bibitem[Br{\`e}s {\em et~al.\/}(2017)Br{\`e}s, Ham, Nichols \&
  Lele]{bres2017unstructured}
{\sc \au{Br{\`e}s, G.~A.}, \au{Ham, F.~E.}, \au{Nichols, J.~W.} \& \au{Lele,
  S.~K.}} \yr{2017}  \at{Unstructured large-eddy simulations of supersonic
  jets}.  \jt{AIAA Journal}  \pg{pp. 1164--1184}.

\bibitem[Br{\`e}s {\em et~al.\/}(2018)Br{\`e}s, Jordan, Jaunet, Le~Rallic,
  Cavalieri, Towne, Lele, Colonius \& Schmidt]{bres2018importance}
{\sc \au{Br{\`e}s, G.~A.}, \au{Jordan, P.}, \au{Jaunet, V.}, \au{Le~Rallic,
  M.}, \au{Cavalieri, A. V.~G.}, \au{Towne, A.}, \au{Lele, S.~K.},
  \au{Colonius, T.} \& \au{Schmidt, O.~T.}} \yr{2018}  \at{Importance of the
  nozzle-exit boundary-layer state in subsonic turbulent jets}.  \jt{Journal of
  Fluid Mechanics}  \bvol{851},  \pg{83--124}.

\bibitem[Br{\`e}s \& Lele(2019)]{bres2019modelling}
{\sc \au{Br{\`e}s, G.~A.} \& \au{Lele, S.~K.}} \yr{2019}  \at{Modelling of jet
  noise: a perspective from large-eddy simulations}.  \jt{Philosophical
  Transactions of the Royal Society A}  \bvol{377}~(2159),  \pg{20190081}.

\bibitem[Brindise {\em et~al.\/}(2017)Brindise, Chiastra, Burzotta, Migliavacca
  \& Vlachos]{brindise2017hemodynamics}
{\sc \au{Brindise, M.~C.}, \au{Chiastra, C.}, \au{Burzotta, F.},
  \au{Migliavacca, F.} \& \au{Vlachos, P.~P.}} \yr{2017}  \at{Hemodynamics of
  stent implantation procedures in coronary bifurcations: An in vitro study}.
  \jt{Annals of biomedical engineering}  \bvol{45}~(3),  \pg{542--553}.

\bibitem[Brindise \& Vlachos(2017)]{brindise2017proper}
{\sc \au{Brindise, M.~C.} \& \au{Vlachos, P.~P.}} \yr{2017}  \at{Proper
  orthogonal decomposition truncation method for data denoising and order
  reduction}.  \jt{Experiments in Fluids}  \bvol{58}~(4),  \pg{28}.

\bibitem[Brown {\em et~al.\/}(1989)Brown, Buchsbaum, Hall, Penhune, Schmitt,
  Watson \& Wyatt]{brown1989observations}
{\sc \au{Brown, E.~D.}, \au{Buchsbaum, S.~B.}, \au{Hall, R.~E.}, \au{Penhune,
  J.~P.}, \au{Schmitt, K.~F.}, \au{Watson, K.~M.} \& \au{Wyatt, D.~C.}}
  \yr{1989}  \at{Observations of a nonlinear solitary wave packet in the kelvin
  wake of a ship}.  \jt{Journal of Fluid Mechanics}  \bvol{204},
  \pg{263--293}.

\bibitem[Camussi(2002)]{camussi2002coherent}
{\sc \au{Camussi, R.}} \yr{2002}  \at{Coherent structure identification from
  wavelet analysis of particle image velocimetry data}.  \jt{Experiments in
  Fluids}  \bvol{32}~(1),  \pg{76--86}.

\bibitem[Camussi \& Guj(1997)]{camussi1997orthonormal}
{\sc \au{Camussi, R.} \& \au{Guj, G.}} \yr{1997}  \at{Orthonormal wavelet
  decomposition of turbulent flows: intermittency and coherent structures}.
  \jt{Journal of Fluid Mechanics}  \bvol{348},  \pg{177--199}.

\bibitem[Charonko {\em et~al.\/}(2010)Charonko, Karri, Schmieg, Prabhu \&
  Vlachos]{charonko2010vitro}
{\sc \au{Charonko, J.}, \au{Karri, S.}, \au{Schmieg, J.}, \au{Prabhu, S.} \&
  \au{Vlachos, P.}} \yr{2010}  \at{In vitro comparison of the effect of stent
  configuration on wall shear stress using time-resolved particle image
  velocimetry}.  \jt{Annals of biomedical engineering}  \bvol{38}~(3),
  \pg{889--902}.

\bibitem[Citriniti \& George(2000)]{citriniti2000reconstruction}
{\sc \au{Citriniti, J.~H.} \& \au{George, W.~K.}} \yr{2000}  \at{Reconstruction
  of the global velocity field in the axisymmetric mixing layer utilizing the
  proper orthogonal decomposition}.  \jt{Journal of Fluid Mechanics}
  \bvol{418},  \pg{137--166}.

\bibitem[Cohen(1995)]{cohen1995time}
{\sc \au{Cohen, L.}} \yr{1995} {\em Time-frequency analysis\/}, ,  \vol{vol.
  778}.  \publ{Prentice hall}.

\bibitem[Delville(1994)]{delville1994characterization}
{\sc \au{Delville, Jo{\"e}l}} \yr{1994}  \at{Characterization of the
  organization in shear layers via the proper orthogonal decomposition}.
  \jt{Applied Scientific Research}  \bvol{53}~(3),  \pg{263--281}.

\bibitem[Discetti {\em et~al.\/}(2013)Discetti, Natale \&
  Astarita]{discetti2013spatial}
{\sc \au{Discetti, S.}, \au{Natale, A.} \& \au{Astarita, T.}} \yr{2013}
  \at{Spatial filtering improved tomographic piv}.  \jt{Experiments in Fluids}
  \bvol{54}~(4),  \pg{1505}.

\bibitem[Farge(1992)]{farge1992wavelet}
{\sc \au{Farge, M.}} \yr{1992}  \at{Wavelet transforms and their applications
  to turbulence}.  \jt{Annual review of fluid mechanics}  \bvol{24}~(1),
  \pg{395--458}.

\bibitem[Farge {\em et~al.\/}(2001)Farge, Pellegrino \&
  Schneider]{farge2001coherent}
{\sc \au{Farge, M.}, \au{Pellegrino, G.} \& \au{Schneider, K.}} \yr{2001}
  \at{Coherent vortex extraction in 3d turbulent flows using orthogonal
  wavelets}.  \jt{Physical Review Letters}  \bvol{87}~(5),  \pg{054501}.

\bibitem[Farge {\em et~al.\/}(1999)Farge, Schneider \& Kevlahan]{farge1999non}
{\sc \au{Farge, M.}, \au{Schneider, K.} \& \au{Kevlahan, N.}} \yr{1999}
  \at{Non-gaussianity and coherent vortex simulation for two-dimensional
  turbulence using an adaptive orthogonal wavelet basis}.  \jt{Physics of
  Fluids}  \bvol{11}~(8),  \pg{2187--2201}.

\bibitem[Fore {\em et~al.\/}(2005)Fore, Tung, Buchanan \&
  Welch]{fore2005nonlinear}
{\sc \au{Fore, L.~B.}, \au{Tung, A.~T.}, \au{Buchanan, J.~R.} \& \au{Welch,
  J.~W.}} \yr{2005}  \at{Nonlinear temporal filtering of time-resolved digital
  particle image velocimetry data}.  \jt{Experiments in Fluids}  \bvol{39}~(1),
   \pg{22--31}.

\bibitem[Freund \& Colonius(2009)]{freund2009turbulence}
{\sc \au{Freund, JB} \& \au{Colonius, T}} \yr{2009}  \at{Turbulence and
  sound-field pod analysis of a turbulent jet}.  \jt{International Journal of
  Aeroacoustics}  \bvol{8}~(4),  \pg{337--354}.

\bibitem[Gamard {\em et~al.\/}(2002)Gamard, George, Jung \&
  Woodward]{gamard2002application}
{\sc \au{Gamard, S.}, \au{George, W.~K.}, \au{Jung, D.} \& \au{Woodward, S.}}
  \yr{2002}  \at{Application of a “slice” proper orthogonal decomposition
  to the far field of an axisymmetric turbulent jet}.  \jt{Physics of Fluids}
  \bvol{14}~(7),  \pg{2515--2522}.

\bibitem[Gamard {\em et~al.\/}(2004)Gamard, Jung \&
  George]{gamard2004downstream}
{\sc \au{Gamard, S.}, \au{Jung, D.} \& \au{George, W.~K.}} \yr{2004}
  \at{Downstream evolution of the most energetic modes in a turbulent
  axisymmetric jet at high reynolds number. part 2. the far-field region}.
  \jt{Journal of Fluid Mechanics}  \bvol{514},  \pg{205--230}.

\bibitem[Ghate {\em et~al.\/}(2020)Ghate, Towne \& Lele]{ghate2020broadband}
{\sc \au{Ghate, A.~S.}, \au{Towne, A.} \& \au{Lele, S.~K.}} \yr{2020}
  \at{Broadband reconstruction of inhomogeneous turbulence using spectral
  proper orthogonal decomposition and gabor modes}.  \jt{Journal of Fluid
  Mechanics}  \bvol{888}.

\bibitem[Glauser \& George(1992)]{glauser1992application}
{\sc \au{Glauser, Mark~N} \& \au{George, William~K}} \yr{1992}  \at{Application
  of multipoint measurements for flow characterization}.  \jt{Experimental
  Thermal and Fluid Science}  \bvol{5}~(5),  \pg{617--632}.

\bibitem[Glauser {\em et~al.\/}(1987)Glauser, Leib \&
  George]{glauser1987coherent}
{\sc \au{Glauser, Mark~N}, \au{Leib, Stewart~J} \& \au{George, William~K}}
  \yr{1987}  \at{Coherent structures in the axisymmetric turbulent jet mixing
  layer}.  \bt{In {\em Turbulent Shear Flows 5\/}},  \pg{pp. 134--145}.
  \publ{Springer}.

\bibitem[Gordeyev \& Thomas(2000)]{gordeyev2000coherent}
{\sc \au{Gordeyev, S.~V.} \& \au{Thomas, F.~O.}} \yr{2000}  \at{Coherent
  structure in the turbulent planar jet. part 1. extraction of proper
  orthogonal decomposition eigenmodes and their self-similarity}.  \jt{Journal
  of Fluid Mechanics}  \bvol{414},  \pg{145--194}.

\bibitem[Gordeyev \& Thomas(2002)]{gordeyev2002coherent}
{\sc \au{Gordeyev, S.~V.} \& \au{Thomas, F.~O.}} \yr{2002}  \at{Coherent
  structure in the turbulent planar jet. part 2. structural topology via pod
  eigenmode projection}.  \jt{Journal of Fluid Mechanics}  \bvol{460},
  \pg{349--380}.

\bibitem[Gu \& Philander(1995)]{gu1995secular}
{\sc \au{Gu, D.} \& \au{Philander, S. G.~H.}} \yr{1995}  \at{Secular changes of
  annual and interannual variability in the tropics during the past century}.
  \jt{Journal of Climate}  \bvol{8}~(4),  \pg{864--876}.

\bibitem[Gudmundsson \& Colonius(2011)]{gudmundsson2011instability}
{\sc \au{Gudmundsson, K.} \& \au{Colonius, T.}} \yr{2011}  \at{Instability wave
  models for the near-field fluctuations of turbulent jets}.  \jt{Journal of
  Fluid Mechanics}  \bvol{689},  \pg{97--128}.

\bibitem[Harris(1978)]{harris1978use}
{\sc \au{Harris, F.~J.}} \yr{1978}  \at{On the use of windows for harmonic
  analysis with the discrete fourier transform}.  \jt{Proceedings of the IEEE}
  \bvol{66}~(1),  \pg{51--83}.

\bibitem[Hellstr{\"o}m {\em et~al.\/}(2015)Hellstr{\"o}m, Ganapathisubramani \&
  Smits]{hellstrom2015evolution}
{\sc \au{Hellstr{\"o}m, L. H.~O.}, \au{Ganapathisubramani, B.} \& \au{Smits,
  A.~J.}} \yr{2015}  \at{The evolution of large-scale motions in turbulent pipe
  flow}.  \jt{Journal of Fluid Mechanics}  \bvol{779},  \pg{701--715}.

\bibitem[Hellstr{\"o}m {\em et~al.\/}(2016)Hellstr{\"o}m, Marusic \&
  Smits]{hellstrom2016self}
{\sc \au{Hellstr{\"o}m, L. H.~O.}, \au{Marusic, I.} \& \au{Smits, A.~J.}}
  \yr{2016}  \at{Self-similarity of the large-scale motions in turbulent pipe
  flow}.  \jt{Journal of Fluid Mechanics}  \bvol{792}.

\bibitem[Hellstr{\"o}m \& Smits(2014)]{hellstrom2014energetic}
{\sc \au{Hellstr{\"o}m, L. H.~O.} \& \au{Smits, A.~J.}} \yr{2014}  \at{The
  energetic motions in turbulent pipe flow}.  \jt{Physics of Fluids}
  \bvol{26}~(12),  \pg{125102}.

\bibitem[Hellstr{\"o}m \& Smits(2017)]{hellstrom2017structure}
{\sc \au{Hellstr{\"o}m, L. H.~O.} \& \au{Smits, A.~J.}} \yr{2017}
  \at{Structure identification in pipe flow using proper orthogonal
  decomposition}.  \jt{Philosophical Transactions of the Royal Society A:
  Mathematical, Physical and Engineering Sciences}  \bvol{375}~(2089),
  \pg{20160086}.

\bibitem[Holmes {\em et~al.\/}(2012)Holmes, Lumley, Berkooz \&
  Rowley]{holmes2012turbulence}
{\sc \au{Holmes, Philip}, \au{Lumley, John~L}, \au{Berkooz, Gahl} \&
  \au{Rowley, Clarence~W}} \yr{2012} {\em Turbulence, coherent structures,
  dynamical systems and symmetry\/}.  \publ{Cambridge university press}.

\bibitem[Huang {\em et~al.\/}(1998)Huang, Shen, Long, Wu, Shih, Zheng, Yen,
  Tung \& Liu]{huang1998empirical}
{\sc \au{Huang, N.~E.}, \au{Shen, Z.}, \au{Long, S.~R}, \au{Wu, M.~C.},
  \au{Shih, H.~H.}, \au{Zheng, Q.}, \au{Yen, N.~C.}, \au{Tung, C.~C.} \&
  \au{Liu, H.~H.}} \yr{1998}  \at{The empirical mode decomposition and the
  hilbert spectrum for nonlinear and non-stationary time series analysis}.
  \jt{Proceedings of the Royal Society of London. Series A: mathematical,
  physical and engineering sciences}  \bvol{454}~(1971),  \pg{903--995}.

\bibitem[Iqbal \& Thomas(2007)]{iqbal2007coherent}
{\sc \au{Iqbal, M.~O.} \& \au{Thomas, F.~O.}} \yr{2007}  \at{Coherent structure
  in a turbulent jet via a vector implementation of the proper orthogonal
  decomposition}.  \jt{Journal of Fluid Mechanics}  \bvol{571},  \pg{281--326}.

\bibitem[Izatt {\em et~al.\/}(1997)Izatt, Kulkarni, Yazdanfar, Barton \&
  Welch]{izatt1997vivo}
{\sc \au{Izatt, J.~A.}, \au{Kulkarni, M.~D.}, \au{Yazdanfar, S.}, \au{Barton,
  J.~K.} \& \au{Welch, A.~J.}} \yr{1997}  \at{In vivo bidirectional color
  doppler flow imaging of picoliter blood volumes using optical coherence
  tomography}.  \jt{Optics letters}  \bvol{22}~(18),  \pg{1439--1441}.

\bibitem[Johansson \& George(2006{\natexlab{{\em a\/}}})]{johansson2006far}
{\sc \au{Johansson, P. B.~V.} \& \au{George, W.~K.}} \yr{2006{\natexlab{{\em
  a\/}}}}  \at{The far downstream evolution of the high-reynolds-number
  axisymmetric wake behind a disk. part 1. single-point statistics}.
  \jt{Journal of Fluid Mechanics}  \bvol{555},  \pg{363--385}.

\bibitem[Johansson \& George(2006{\natexlab{{\em b\/}}})]{johansson2006far2}
{\sc \au{Johansson, P. B.~V.} \& \au{George, W.~K.}} \yr{2006{\natexlab{{\em
  b\/}}}}  \at{The far downstream evolution of the high-reynolds-number
  axisymmetric wake behind a disk. part 2. slice proper orthogonal
  decomposition}.  \jt{Journal of Fluid Mechanics}  \bvol{555},  \pg{387--408}.

\bibitem[Johansson {\em et~al.\/}(2002)Johansson, George \&
  Woodward]{johansson2002proper}
{\sc \au{Johansson, P. B.~V.}, \au{George, W.~K.} \& \au{Woodward, S.~H.}}
  \yr{2002}  \at{Proper orthogonal decomposition of an axisymmetric turbulent
  wake behind a disk}.  \jt{Physics of Fluids}  \bvol{14}~(7),
  \pg{2508--2514}.

\bibitem[Jung {\em et~al.\/}(2004)Jung, Gamard \& George]{jung2004downstream}
{\sc \au{Jung, D.}, \au{Gamard, S.} \& \au{George, W.~K.}} \yr{2004}
  \at{Downstream evolution of the most energetic modes in a turbulent
  axisymmetric jet at high reynolds number. part 1. the near-field region}.
  \jt{Journal of Fluid Mechanics}  \bvol{514},  \pg{173--204}.

\bibitem[Khaliq {\em et~al.\/}(2009)Khaliq, Ouarda, Gachon, Sushama \&
  St-Hilaire]{khaliq2009identification}
{\sc \au{Khaliq, M.~N.}, \au{Ouarda, T. B. M.~J.}, \au{Gachon, P.},
  \au{Sushama, L.} \& \au{St-Hilaire, A.}} \yr{2009}  \at{Identification of
  hydrological trends in the presence of serial and cross correlations: A
  review of selected methods and their application to annual flow regimes of
  canadian rivers}.  \jt{Journal of Hydrology}  \bvol{368}~(1-4),
  \pg{117--130}.

\bibitem[Kostas {\em et~al.\/}(2005)Kostas, Soria \&
  Chong]{kostas2005comparison}
{\sc \au{Kostas, J.}, \au{Soria, J.} \& \au{Chong, M.~S.}} \yr{2005}  \at{A
  comparison between snapshot pod analysis of piv velocity and vorticity data}.
   \jt{Experiments in Fluids}  \bvol{38}~(2),  \pg{146--160}.

\bibitem[Lacombe {\em et~al.\/}(2012)Lacombe, McCartney \&
  Forkuor]{lacombe2012drying}
{\sc \au{Lacombe, G.}, \au{McCartney, M.} \& \au{Forkuor, G.}} \yr{2012}
  \at{Drying climate in ghana over the period 1960--2005: evidence from the
  resampling-based mann-kendall test at local and regional levels}.
  \jt{Hydrological Sciences Journal}  \bvol{57}~(8),  \pg{1594--1609}.

\bibitem[Lumley(1967)]{lumley1967structure}
{\sc \au{Lumley, J.~L.}} \yr{1967}  \at{The structure of inhomogeneous
  turbulent flows}.  \jt{Atmospheric turbulence and radio wave propagation} .

\bibitem[Lumley(1970)]{lumley1970stochastic}
{\sc \au{Lumley, J.~L.}} \yr{1970} {\em Stochastic tools in turbulence\/}.
  \publ{Courier Corporation}.

\bibitem[Meneveau {\em et~al.\/}(1992)Meneveau, Lund \&
  Moin]{meneveau1992search}
{\sc \au{Meneveau, C}, \au{Lund, TS} \& \au{Moin, Parviz}} \yr{1992} Search for
  subgrid scale parameterization by projection pursuit regression.  \bt{In {\em
  Proceedings of Summer Program\/}},  \pg{pp. 61--81}. Stanford University.

\bibitem[Meyers {\em et~al.\/}(1993)Meyers, Kelly \&
  O'Brien]{meyers1993introduction}
{\sc \au{Meyers, S.~D.}, \au{Kelly, B.~G.} \& \au{O'Brien, J.~J.}} \yr{1993}
  \at{An introduction to wavelet analysis in oceanography and meteorology: With
  application to the dispersion of yanai waves}.  \jt{Monthly weather review}
  \bvol{121}~(10),  \pg{2858--2866}.

\bibitem[Muralidhar {\em et~al.\/}(2019)Muralidhar, Podvin, Mathelin \&
  Fraigneau]{muralidhar2019spatio}
{\sc \au{Muralidhar, S.~D.}, \au{Podvin, B.}, \au{Mathelin, L.} \&
  \au{Fraigneau, Y.}} \yr{2019}  \at{Spatio-temporal proper orthogonal
  decomposition of turbulent channel flow}.  \jt{Journal of Fluid Mechanics}
  \bvol{864},  \pg{614--639}.

\bibitem[Nekkanti \& Schmidt(2020)]{nekkanti2020modal}
{\sc \au{Nekkanti, A.} \& \au{Schmidt, O.~T.}} \yr{2020}  \at{Modal analysis of
  acoustic directivity in turbulent jets}.  \jt{AIAA Journal}  \pg{pp. 1--12}.

\bibitem[Nidhan {\em et~al.\/}(2020)Nidhan, Chongsiripinyo, Schmidt \&
  Sarkar]{nidhan2020spectral}
{\sc \au{Nidhan, S}, \au{Chongsiripinyo, K}, \au{Schmidt, OT} \& \au{Sarkar,
  S}} \yr{2020}  \at{Spectral proper orthogonal decomposition analysis of the
  turbulent wake of a disk at re= 50 000}.  \jt{Physical Review Fluids}
  \bvol{5}~(12),  \pg{124606}.

\bibitem[Onorato {\em et~al.\/}(2000)Onorato, Camussi \&
  Iuso]{onorato2000small}
{\sc \au{Onorato, M.}, \au{Camussi, R.} \& \au{Iuso, G.}} \yr{2000}  \at{Small
  scale intermittency and bursting in a turbulent channel flow}.  \jt{Physical
  Review E}  \bvol{61}~(2),  \pg{1447}.

\bibitem[Picard \& Delville(2000)]{picard2000pressure}
{\sc \au{Picard, C} \& \au{Delville, J}} \yr{2000}  \at{Pressure velocity
  coupling in a subsonic round jet}.  \jt{International Journal of Heat and
  Fluid Flow}  \bvol{21}~(3),  \pg{359--364}.

\bibitem[Pickering {\em et~al.\/}(2019)Pickering, Rigas, Nogueira, Cavalieri,
  Schmidt \& Colonius]{pickering2019lift}
{\sc \au{Pickering, E.}, \au{Rigas, G.}, \au{Nogueira, P. A.~S.},
  \au{Cavalieri, A. V.~G.}, \au{Schmidt, O.~T.} \& \au{Colonius, T.}} \yr{2019}
   \at{Lift-up, kelvin-helmholtz and orr mechanisms in turbulent jets}.
  \jt{arXiv preprint arXiv:1909.09737} .

\bibitem[Raiola {\em et~al.\/}(2015)Raiola, Discetti \& Ianiro]{raiola2015piv}
{\sc \au{Raiola, M.}, \au{Discetti, S.} \& \au{Ianiro, A.}} \yr{2015}  \at{On
  piv random error minimization with optimal pod-based low-order
  reconstruction}.  \jt{Experiments in Fluids}  \bvol{56}~(4),  \pg{75}.

\bibitem[Rodionov(2006)]{rodionov2006use}
{\sc \au{Rodionov, S.~N.}} \yr{2006}  \at{Use of prewhitening in climate regime
  shift detection}.  \jt{Geophysical Research Letters}  \bvol{33}~(12).

\bibitem[Rowley \& Dawson(2017)]{rowley2017model}
{\sc \au{Rowley, C.~W.} \& \au{Dawson, S. T.~M.}} \yr{2017}  \at{Model
  reduction for flow analysis and control}.  \jt{Annual Review of Fluid
  Mechanics}  \bvol{49},  \pg{387--417}.

\bibitem[Rowley {\em et~al.\/}(2009)Rowley, Mezi{\'c}, Bagheri, Schlatter,
  Henningson {\em et~al.\/}]{rowley2009spectral}
{\sc \au{Rowley, C.~W.}, \au{Mezi{\'c}, I.}, \au{Bagheri, S.}, \au{Schlatter,
  P.}, \au{Henningson, D.} \& \au{others}} \yr{2009}  \at{Spectral analysis of
  nonlinear flows}.  \jt{Journal of Fluid Mechanics}  \bvol{641}~(1),
  \pg{115--127}.

\bibitem[Samimy {\em et~al.\/}(2007)Samimy, Debiasi, Caraballo, Serrani, Yuan,
  Little \& Myatt]{samimy2007feedback}
{\sc \au{Samimy, M.}, \au{Debiasi, M.}, \au{Caraballo, E.}, \au{Serrani, A.},
  \au{Yuan, X.}, \au{Little, J.} \& \au{Myatt, J.~H.}} \yr{2007}  \at{Feedback
  control of subsonic cavity flows using reduced-order models}.  \jt{Journal of
  Fluid Mechanics}  \bvol{579},  \pg{315}.

\bibitem[Schmid(2010)]{schmid2010dynamic}
{\sc \au{Schmid, P.~J.}} \yr{2010}  \at{Dynamic mode decomposition of numerical
  and experimental data}.  \jt{Journal of Fluid Mechanics}  \bvol{656},
  \pg{5--28}.

\bibitem[Schmidt \& Colonius(2020)]{schmidt2020guide}
{\sc \au{Schmidt, O.~T.} \& \au{Colonius, T.}} \yr{2020}  \at{Guide to spectral
  proper orthogonal decomposition}.  \jt{AIAA Journal}  \pg{pp. 1--11}.

\bibitem[Schmidt {\em et~al.\/}(2017{\natexlab{{\em a\/}}})Schmidt, Colonius \&
  Br{\'e}s]{schmidt2017wavepacket}
{\sc \au{Schmidt, O.~T.}, \au{Colonius, T.} \& \au{Br{\'e}s, G.~A.}}
  \yr{2017{\natexlab{{\em a\/}}}} Wavepacket intermittency and its role in
  turbulent jet noise.  \bt{In {\em 55th AIAA Aerospace Sciences Meeting\/}},
  \pg{p. 0686}.

\bibitem[Schmidt {\em et~al.\/}(2017{\natexlab{{\em b\/}}})Schmidt, Towne,
  Colonius, Cavalieri, Jordan \& Br{\`e}s]{schmidt2017wavepackets}
{\sc \au{Schmidt, O.~T.}, \au{Towne, A.}, \au{Colonius, T.}, \au{Cavalieri, A.
  V.~G.}, \au{Jordan, P.} \& \au{Br{\`e}s, G.~A.}} \yr{2017{\natexlab{{\em
  b\/}}}}  \at{Wavepackets and trapped acoustic modes in a turbulent jet:
  coherent structure eduction and global stability}.  \jt{Journal of Fluid
  Mechanics}  \bvol{825},  \pg{1153--1181}.

\bibitem[Schmidt {\em et~al.\/}(2018)Schmidt, Towne, Rigas, Colonius \&
  Br{\`e}s]{schmidt2018spectral}
{\sc \au{Schmidt, O.~T.}, \au{Towne, A.}, \au{Rigas, G.}, \au{Colonius, T.} \&
  \au{Br{\`e}s, G.~A.}} \yr{2018}  \at{Spectral analysis of jet turbulence}.
  \jt{Journal of Fluid Mechanics}  \bvol{855},  \pg{953--982}.

\bibitem[Sciacchitano \& Scarano(2014)]{sciacchitano2014elimination}
{\sc \au{Sciacchitano, A.} \& \au{Scarano, F.}} \yr{2014}  \at{Elimination of
  piv light reflections via a temporal high pass filter}.  \jt{Measurement
  Science and Technology}  \bvol{25}~(8),  \pg{084009}.

\bibitem[Serinaldi \& Kilsby(2015)]{serinaldi2015stationarity}
{\sc \au{Serinaldi, F.} \& \au{Kilsby, C.~G.}} \yr{2015}  \at{Stationarity is
  undead: Uncertainty dominates the distribution of extremes}.  \jt{Advances in
  Water Resources}  \bvol{77},  \pg{17--36}.

\bibitem[Sirovich(1987)]{sirovich1987turbulence}
{\sc \au{Sirovich, L.}} \yr{1987}  \at{Turbulence and the dynamics of coherent
  structures. i. coherent structures}.  \jt{Quarterly of applied mathematics}
  \bvol{45}~(3),  \pg{561--571}.

\bibitem[Son \& Kihm(2001)]{son2001evaluation}
{\sc \au{Son, S.~Y.} \& \au{Kihm, K.~D.}} \yr{2001}  \at{Evaluation of
  transient turbulent flow fields using digital cinematographic particle image
  velocimetry}.  \jt{Experiments in Fluids}  \bvol{30}~(5),  \pg{537--550}.

\bibitem[Stewart \& Vlachos(2012)]{stewart2012vortex}
{\sc \au{Stewart, K.~C.} \& \au{Vlachos, P.~P.}} \yr{2012}  \at{Vortex rings in
  radially confined domains}.  \jt{Experiments in Fluids}  \bvol{53}~(4),
  \pg{1033--1044}.

\bibitem[Stockwell {\em et~al.\/}(1996)Stockwell, Mansinha \&
  Lowe]{stockwell1996localization}
{\sc \au{Stockwell, R.~G.}, \au{Mansinha, L.} \& \au{Lowe, R.~P.}} \yr{1996}
  \at{Localization of the complex spectrum: the s transform}.  \jt{IEEE
  transactions on signal processing}  \bvol{44}~(4),  \pg{998--1001}.

\bibitem[von Storch(1995)]{von1995misuses}
{\sc \au{von Storch, H.}} \yr{1995}  \at{Misuses of statistical analysis in
  climate research. in: Analysis of climate variability. springer, berlin,
  heidelberg.}  \pg{pp. 11--26}.

\bibitem[Suzuki \& Colonius(2006)]{suzuki2006instability}
{\sc \au{Suzuki, T.} \& \au{Colonius, T.}} \yr{2006}  \at{Instability waves in
  a subsonic round jet detected using a near-field phased microphone array}.
  \jt{Journal of Fluid Mechanics}  \bvol{565},  \pg{197--226}.

\bibitem[Taira {\em et~al.\/}(2017)Taira, Brunton, Dawson, Rowley, Colonius,
  McKeon, Schmidt, Gordeyev, Theofilis \& Ukeiley]{taira2017modal}
{\sc \au{Taira, K.}, \au{Brunton, S.~L.}, \au{Dawson, S. T.~M.}, \au{Rowley,
  C.~W.}, \au{Colonius, T.}, \au{McKeon, B.~J.}, \au{Schmidt, O.~T.},
  \au{Gordeyev, S.}, \au{Theofilis, V.} \& \au{Ukeiley, L.~S.}} \yr{2017}
  \at{Modal analysis of fluid flows: An overview}.  \jt{AIAA Journal}  \pg{pp.
  4013--4041}.

\bibitem[Tinney {\em et~al.\/}(2008{\natexlab{{\em a\/}}})Tinney, Glauser \&
  Ukeiley]{tinney2008low}
{\sc \au{Tinney, C.~E.}, \au{Glauser, M.~N.} \& \au{Ukeiley, L.~S.}}
  \yr{2008{\natexlab{{\em a\/}}}}  \at{Low-dimensional characteristics of a
  transonic jet. part 1. proper orthogonal decomposition}.  \jt{Journal of
  Fluid Mechanics}  \bvol{612},  \pg{107--141}.

\bibitem[Tinney {\em et~al.\/}(2008{\natexlab{{\em b\/}}})Tinney, Ukeiley \&
  Glauser]{tinney2008low2}
{\sc \au{Tinney, C.~E.}, \au{Ukeiley, L.~S.} \& \au{Glauser, M.~N.}}
  \yr{2008{\natexlab{{\em b\/}}}}  \at{Low-dimensional characteristics of a
  transonic jet. part 2. estimate and far-field prediction}.  \jt{Journal of
  Fluid Mechanics}  \bvol{615},  \pg{53--92}.

\bibitem[Tissot {\em et~al.\/}(2017)Tissot, Laj{\'u}s~Jr, Cavalieri \&
  Jordan]{tissot2017wave}
{\sc \au{Tissot, Gilles}, \au{Laj{\'u}s~Jr, Francisco~C}, \au{Cavalieri,
  Andr{\'e}~VG} \& \au{Jordan, Peter}} \yr{2017}  \at{Wave packets and orr
  mechanism in turbulent jets}.  \jt{Physical Review Fluids}  \bvol{2}~(9),
  \pg{093901}.

\bibitem[Towne {\em et~al.\/}(2017)Towne, Cavalieri, Jordan, Colonius, Jaunet
  \& Br{\`e}s]{towne2017acoustic}
{\sc \au{Towne, A.}, \au{Cavalieri, A. V.~G.}, \au{Jordan, P.}, \au{Colonius,
  T.and~Schmidt, O.}, \au{Jaunet, V.} \& \au{Br{\`e}s, G.~A.}} \yr{2017}
  \at{Acoustic resonance in the potential core of subsonic jets}.  \jt{Journal
  of Fluid Mechanics}  \bvol{825},  \pg{1113--1152}.

\bibitem[Towne \& Liu(2019)]{towne2019time}
{\sc \au{Towne, A.} \& \au{Liu, P.}} \yr{2019}  \at{Time-frequency analysis of
  intermittent coherent structures in turbulent flows}.  \jt{Bulletin of the
  American Physical Society}  \bvol{64}.

\bibitem[Towne {\em et~al.\/}(2018)Towne, Schmidt \&
  Colonius]{towne2018spectral}
{\sc \au{Towne, A.}, \au{Schmidt, O.~T.} \& \au{Colonius, T.}} \yr{2018}
  \at{Spectral proper orthogonal decomposition and its relationship to dynamic
  mode decomposition and resolvent analysis}.  \jt{Journal of Fluid Mechanics}
  \bvol{847},  \pg{821--867}.

\bibitem[Tutkun {\em et~al.\/}(2008)Tutkun, Johansson \&
  George]{tutkun2008three}
{\sc \au{Tutkun, M.}, \au{Johansson, P. B.~V.} \& \au{George, W.~K}} \yr{2008}
  \at{Three-component vectorial proper orthogonal decomposition of axisymmetric
  wake behind a disk}.  \jt{AIAA Journal}  \bvol{46}~(5),  \pg{1118--1134}.

\bibitem[V{\'e}tel {\em et~al.\/}(2011)V{\'e}tel, Garon \&
  Pelletier]{vetel2011denoising}
{\sc \au{V{\'e}tel, J.}, \au{Garon, A.} \& \au{Pelletier, D.}} \yr{2011}
  \at{Denoising methods for time-resolved piv measurements}.  \jt{Experiments
  in Fluids}  \bvol{51}~(4),  \pg{893--916}.

\bibitem[Welch(1967)]{welch1967use}
{\sc \au{Welch, P.}} \yr{1967}  \at{The use of fast fourier transform for the
  estimation of power spectra: a method based on time averaging over short,
  modified periodograms}.  \jt{IEEE Transactions on audio and electroacoustics}
   \bvol{15}~(2),  \pg{70--73}.

\bibitem[Zhang {\em et~al.\/}(2001)Zhang, Harvey, Hogg \&
  Yuzyk]{zhang2001trends}
{\sc \au{Zhang, X.}, \au{Harvey, K.~D.}, \au{Hogg, W.~D.} \& \au{Yuzyk, T.~R.}}
  \yr{2001}  \at{Trends in canadian streamflow}.  \jt{Water Resources Research}
   \bvol{37}~(4),  \pg{987--998}.

\bibitem[Zhang {\em et~al.\/}(2003)Zhang, Guo, Wang, He, Lee \&
  Loew]{zhang2003comparison}
{\sc \au{Zhang, Y.}, \au{Guo, Z.}, \au{Wang, W.}, \au{He, S.}, \au{Lee, T.} \&
  \au{Loew, M.}} \yr{2003}  \at{A comparison of the wavelet and short-time
  fourier transforms for doppler spectral analysis}.  \jt{Medical engineering
  \& physics}  \bvol{25}~(7),  \pg{547--557}.

\end{thebibliography}
\end{document}